\documentclass[twocolumn,preprintnumbers,superscriptaddress,amsmath,amssymb]{revtex4-1}
\usepackage{float}
\usepackage{graphicx}
\usepackage{dcolumn}
\usepackage{bm}
\usepackage{epstopdf}
\usepackage{xcolor}
\usepackage{textcomp}
\usepackage{soul}
\usepackage{csquotes}
\usepackage{amsbsy}
\usepackage{mathrsfs} 
\usepackage{amsmath}
\usepackage{bigints} 
\usepackage{amsthm,amssymb} 
\usepackage[utf8]{inputenc}
\usepackage[english]{babel}
\usepackage[shortlabels]{enumitem}
\usepackage{float}
\usepackage{mathrsfs,bigints,mathtools,dsfont}
\usepackage[colorlinks=true,linkcolor=blue,citecolor=blue]{hyperref}%
\usepackage[toc]{appendix}
\newcommand\norm[1]{\left\lVert#1\right\rVert}

\newtheorem{theorem}{Theorem}

\newtheorem{lemma}[theorem]{Lemma}

\begin{document}
\title{{Stability analysis of intralayer synchronization in time-varying multilayer networks with generic coupling functions}}
\author{Md Sayeed Anwar}\affiliation{Physics and Applied Mathematics Unit, Indian Statistical Institute, 203 B. T. Road, Kolkata 700108, India}
\author{Sarbendu Rakshit}\affiliation{Physics and Applied Mathematics Unit, Indian Statistical Institute, 203 B. T. Road, Kolkata 700108, India}
\author{Dibakar Ghosh}\email{dibakar@isical.ac.in}\affiliation{Physics and Applied Mathematics Unit, Indian Statistical Institute, 203 B. T. Road, Kolkata 700108, India}
\author{Erik M. Bollt}\affiliation{Department of Mathematics, Department of Electrical and Computer Engineering, Department of Physics, Clarkson University, Potsdam, New York 13699}

\begin{abstract}
\par The stability analysis of synchronization patterns on generalized network structures is of immense importance nowadays. In this article, we scrutinize the stability of intralayer synchronous state in temporal multilayer hypernetworks, where each dynamic units in a layer communicate with others through various independent time-varying connection mechanisms. Here, dynamical units within and between layers may be interconnected through arbitrary generic coupling functions. We show that intralayer synchronous state exists as an invariant solution. Using fast switching stability criteria, we derive the condition for stable coherent state  in terms of associated time-averaged network structure, and in some instances we are able to separate the transverse subspace optimally. Using simultaneous block diagonalization of coupling matrices, we derive the synchronization stability condition without considering time-averaged network structure. Finally, we verify our {analytically derived results} through a series of numerical simulations on synthetic and real-world neuronal networked systems.
\end{abstract}

\maketitle

\section{Introduction} 
\par In the last decade, the study on multilayer networks \cite{boccaletti2014structure,bianconi2018multilayer} has become a {prosperous} research area due to its ability in describing many real-world systems \cite{mobility,power,air,social,ecology} with heterogeneous interacting layers. Another interesting generalized  network structure is hypernetworks \cite{sorrentino_njp,sorrentino2007network,DBLP:conf/iclr/HaDL17} where an ensemble of nodes are simultaneously interconnected through various independent connection topologies. For instance, in neuronal network, the neurons are interconnected via chemical and electrical pathways \cite{rakshit2018emergence}. The precise illustrative power of these generalized structures naturally comes up with increasing analytical complication to scrutinize stability of collective behaviors emerge on them. Synchronization \cite{synchronization1,synchronization2,synchronization3,synchronization4,skardal2014optimal}, one such collective phenomena whose stability analysis is very significant as it generally does not emerge if this is not stable.

\par To execute systematic stability analysis of {synchronization} state in large networks \cite{restrepo2006synchronization,restrepo2005onset}, the crucial step is to split up whole space of the variational equation into parallel and transverse subspaces, and evaluate the maximum transverse Lyapunov exponent to find out if perturbations associated with transverse subspace will die out or not. For stability of complete synchronization in large scale network of identical oscillators, the master stability function (MSF) \cite{pecora1998master} approach provides the essential tool, which further broaden in various ways \cite{sun2009master,restrepo2005onset,nishikawa2006synchronization}. In case of hypernetworks, the stability of synchronization has been analyzed by dimensionality reduction \cite{sorrentino_njp} of master stability equation through simultaneous block diagonalization of coupling matrices \cite{sbd_sorrentino}. To scrutinize stability of coherent state in multilayer networks a few new procedures has been provided recently \cite{belykh2019synchronization,multiplex_decom,rakshit2021relay}. However, all of these studies have been performed for static network structures, while most of the real-world systems \cite{onnela2007structure,pastor2004evolution,skufca2004communication,wasserman1994social,motter2013spontaneous} are time-varying \cite{ghosh2022synchronized}, meaning that the connection topologies between generic agents vary over time \cite{holme2012temporal,taylor2010spontaneous}. In {Ref.} \cite{stilwell2006sufficient}, the authors analyzed stability of synchronous state in temporal networks through fast switching stability procedure. Other than fast switching approach, the stability of coherent state in time-varying networks can be performed by connection graph stability method \cite{belykh2004blinking,belykh2004connection}. Very recently, Zhang et al. \cite{zhang2020symmetry} have proposed an all-round simultaneous block diagonalization framework that enables the stability analysis of synchronization patterns for monolayer, multilayer, temporal network structures \cite{zhang2021unified}. The main theme of present work is to analyze the stability of synchronization in generic temporal multilayer hypernetworks, particularly intralayer synchronization \cite{intra1,rakshit2020intralayer}. Until now, the study of synchronization on temporal multilayer networks have been done mainly on case-by-case basis, either by taking multiplex network structure \cite{rakshit2018synchronization,rakshit2020intralayer}, or with assumption of certain type of intralayer, and interlayer coupling functions (mostly diffusive) \cite{rakshit2017time,rakshit2020intralayer}.

\par Here we propose a general mathematical framework to investigate intralayer synchronization phenomena on temporal multilayer hypernetwork. In particular, we consider a group of generic units, represented by nodes of the multilayer hypernetwork, interacting within and between the layers through completely arbitrary linear or nonlinear coupling functions with only consideration of static interlayer links. In such broad circumstances, we uncover that the intralayer synchronous state emerges as a stable invariant solution. Besides, we derive the necessary condition for emergence of stable intralayer coherent state using fast switching stability method and simultaneous block diagonalization technique. We show that the temporal multilayer hypernetwork achieves stable intralayer synchronization state for adequately fast switching whenever the corresponding time-averaged network does. Further, we show the master stability equation optimally decouples into independent transverse modes if one time-average Laplacian matrix commutes with all other time-average Laplacians and interlayer adjacencies. On the other hand, through the simultaneous block diagonalization technique, we show that the parallel and transverse modes can be separated without considering time-average network structure, and furthermore the dimensionality reduction of master stability equation can be possible even if the coupling matrices are not commutative. Lastly, all of our analytical results are verified with the numerical simulations of paradigmatic and real-world neuronal networked systems.

\par The rest of this article is organized as follows. In Sec. \ref{model}, we propose the generalized model for time-varying multilayer hypernetwork. The analytical results associated with intralayer synchronization state is illustrated in Sec. \ref{Analytical Results}. In Sec. \ref{invariance}, the invariance condition for the coherent state is derived. Subsequently, in Sec. \ref{Stability}, we establish the condition for stability of synchronous solution. The corresponding numerical results are provided in the subsequent Sec. \ref{numerical}. Finally, we sum up all our results and conclude in Sec. \ref{conclusion}.

\section{Generalized Mathematical Model: Multilayer Hypernetworks}\label{model}
\par We consider a multilayer network consists of two layers, each composed of $N$ generic units. In each layer, the nodes are simultaneously interconnected through two or more topologically different connections, called tiers. In particular, we assume $N$ nodes in each layer are collaborated via $M$ distinct tiers characterizing various types of connections between the dynamical units. We assume that the equations of motion defining the dynamics of our multilayer hypernetwork can be represented by the following sets of equations,

\begin{equation}\label{eq.1}
\begin{array}{lcl}
\dot{\bf x}_i(t) = F_1({\bf x}_i)+\sum\limits_{\beta=1}^{M}\epsilon_\beta\sum\limits_{j=1}^{N}{\mathscr{A}_{ij}^{[1,\beta]}}(t) G_{\beta}^{[1]}({\bf x}_i,{\bf x}_j) \\ ~~~~~~~~~~~~~~~~~~~~ + \lambda \sum\limits_{j=1}^{N}{\mathscr{B}_{ij}^{[1]}}H_1({\bf x}_i,{\bf y}_j), \\
\dot{\bf y}_i(t) = F_2({\bf y}_i)+\sum\limits_{\beta=1}^{M}\epsilon_\beta\sum\limits_{j=1}^{N}{\mathscr{A}_{ij}^{[2,\beta]}}(t) G_{\beta}^{[2]}({\bf y}_i,{\bf y}_j) \\~~~~~~~~~~~~~~~~~~~~ + \lambda \sum\limits_{j=1}^{N}{\mathscr{B}_{ij}^{[2]}}H_2({\bf y}_i,{\bf x}_j), \hspace{10 pt} i=1,2,...,N.
\end{array}
\end{equation}
Here $\mathbf{x}_i(t)$ and $\mathbf{y}_i(t)$ are the $d$-dimensional state vectors illustrating the dynamics of generic node $i$, $F_{1,2} : \mathbb{R}^d \rightarrow  \mathbb{R}^d$ represent the isolated node dynamics presumed to be identical for each generic agent of a particular layer, but different for different layers.  $G_{\beta}^{[l]}  : \mathbb{R}^d\times \mathbb{R}^d \rightarrow  \mathbb{R}^d$ and $H_{1,2} : \mathbb{R}^d\times \mathbb{R}^d \rightarrow  \mathbb{R}^d$ are continuously differentiable functions, representing the  output vectorial functions within each layer for tier-$\beta$ and  between the layers, respectively. Here, we assume the coupling functions to be arbitrary nonlinear functions. Further, the intralayer coupling functions conforming to tier-$\beta$ can be different for different layers and the interlayer functions are non-symmetric, i.e.,  $H_l(x, y)$ is different from $H_l(y, x)$, $l=1,2$. The real-valued parameters $\epsilon_ \beta$ and $\lambda$ describe the intralayer coupling strength for the tier-$\beta$ and interlayer coupling strength, respectively.

\par Moreover, the adjacency matrix $\mathscr{A}^ {[l, \beta]}(t)$ recounts the intralayer network structure associated with the tier-$\beta$ of layer-$l$ at a time instance $t$, where $\mathscr{A}_{ij}^{[l,\beta ]} (t) = 1$ if the $i^{\mbox{th}}$ and $j^{\mbox{th}}$ nodes are interconnected in tier-$\beta$ of layer-$l$ at time t and zero otherwise. The layer-wise connections of the multilayer hypernetwork are varying in time through the probabilistic modification of the whole intralayer network with a rewiring frequency $f$. Smaller $f$ indicates that the intralayer links are almost static, while large $f$ corresponds to very fast swapping of the links over time. Here, we assume that the network structure of tier-$\beta$ for both the layers are equivalent. Although, their corresponding adjacency matrices will not usually be the same because of the {kinematic} behavior of the links of each individual tier. The adjacency matrices $\mathscr{A}^{[l,\beta ]}(t)$ define {the associated} zero-row sum Laplacian matrices {$\mathscr{L}^ {[l, \beta]}(t)$}, whose diagonal entries {are $\mathscr{L}_{ii}^{[l,\beta ]} (t)=\sum_{j=1}^{N}\mathscr{A}_{ij}^{[l,\beta ]}(t)$,} and off-diagonal entities are obtained by just considering negatives of the non-diagonal entries in $\mathscr{A}^{[l,\beta ]}(t)${, i.e., $\mathscr{L}_{ij}^{[l,\beta ]}(t)=-\mathscr{A}_{ij}^{[l,\beta ]}(t)$ for $i\ne j$}. On the other hand, the interlayer connections between the nodes of two layers are characterized by the interlayer adjacency {matrices} $\mathscr{B}^{[l]}$, $(l = 1, 2)$. $\mathscr{B}_{ij}^{[l]}= 1 $ when there is a pairwise link between the $i^{\mbox{th}}$ node of one layer and the $j^{\mbox{th}}$ node of the other layer, and zero otherwise. The interlayer links are {considered to be} static over time. We define the intralayer degree of the $i^{\mbox{th}}$ node of tier-$\beta$ in layer-$l$ as $d_i^{[l,\beta]}(t)=\sum_{j=1}^{N}\mathscr{A}_{ij}^{[l,\beta]}(t)$ and the interlayer node degree in layer-$l$ is denoted by $e_i^{[l]} =\sum_{j=1}^{N}\mathscr{B}_{ij}^{[l]}$. This is the most comprehensive system for a temporal multilayer hypernetwork we can consider, as the coupling functions and the adjacency matrices does not fall under extra limitations.

\begin{figure}[ht]
\centerline{\includegraphics[scale=0.225]{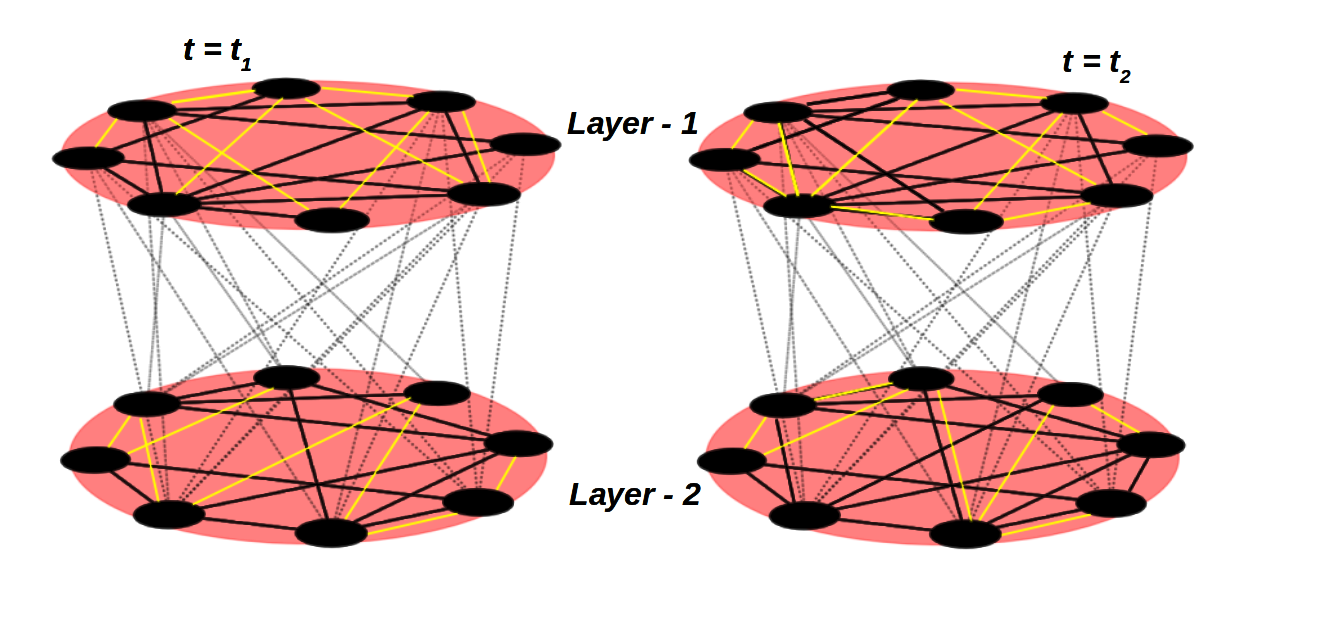}}
\caption{{\bf Schematic representation of temporal multilayer hypernetwork.} The left and right panel, respectively describes the structure of multilayer hypernetwork at two time instances $t=t_1$ and $t=t_2$. Solid black and yellow lines in each layer represents the interconnection between nodes through two distinct tiers. Dotted black lines represents static interlayer connections.}\label{fig.1}
\end{figure}
	
\par Figure \ref{fig.1} represents a schematic sketch of temporal multilayer hypernetwork consisting of two layers, each composed of $N=8$ units and $M=2$ distinct tiers. The connection between nodes through different tiers are shown in two different colors, solid black lines corresponds to one tier, and solid yellow for another one. The left and right panels delineates the structure of the multilayer network for two different time instances $t=t_1$ and $t=t_2$. Here, the connections between nodes of two different layers are static over time, which are shown in dotted black line. 

\par Throughout the following section, our primary target is to derive analytically the necessary conditions for which the intralayer coherent state of the multilayer hypernetwork \eqref{eq.1} is stable.

\section{Analytical Results}\label{Analytical Results}
\subsection{Invariance of Intralayer Synchronization }\label{invariance}
\par Intralayer synchronization state emerges in the multilayer network \eqref{eq.1} if each node in individual layers evolves synchronously with other nodes of the same layer. Mathematically, there exists two solutions $\mathbf{x}_s(t),\mathbf{y}_s(t) \in \mathbb{R}^d$ such that, 
\begin{equation*}
\begin{array}{l}
\norm{\mathbf{x}_i(t)-\mathbf{x}_s(t)} \to 0, ~~ \mbox{and} \\  \norm{\mathbf{y}_i(t)-\mathbf{y}_s(t)} \to 0 ~~ \mbox{as}~~ t \to 0 ~~ \mbox{for} ~~ i=1,2,\dots,N.
\end{array}		
\end{equation*} 
Then the associated synchronization manifold is defined by, 
\begin{equation*}
\begin{array}{l}
\mathcal{S}=\{(\mathbf{x}_s(t),\mathbf{y}_s(t)) \subset \mathbb{R}^{2d} : \mathbf{x}_i(t)=\mathbf{x}_s(t), ~~ \mbox{and} \\ \mathbf{y}_i(t)=\mathbf{y}_s(t) ~~ \mbox{for} ~~ i=1,2,\dots,N ~~ \mbox{and} ~~ t \in \mathbb{R}^+ \}.	
\end{array} 
\end{equation*} 
If the coupling functions are confined to synchronization solutions, i.e., 
\begin{equation*}
	\begin{array}{l}
	G_{\beta}^{[1]}({\bf x}_s,{\bf x}_s)=0, G_{\beta}^{[2]}({\bf y}_s,{\bf y}_s)=0, \; \mbox{and}\; \\ H_1({\bf x}_s,{\bf y}_s)=0,  H_2({\bf y}_s,{\bf x}_s)=0,	
	\end{array}
\end{equation*} 
then the existence and invariance of an intralayer coherent state will be guaranteed. Furthermore we consider arbitrary sort of coupling functions. In this case, as the network structure varies with respect to time, the existence and invariance condition on the network topologies for the emergence of intralayer synchronization is a challenging problem.

\par Suppose the multilayer hypernetwork begins to evolve with the intralayer coherence solution at a time instance $t = t_0$, then $\mathbf{x}_i(t_{0})=\mathbf{x}_s{(t_{0})}$ and $\mathbf{y}_i(t_{0})=\mathbf{y}_s{(t_{0})}$, for $i=1,2,\dots,N$. Then the evolution of the $i^{\mbox{th}}$ node for the two layers at $t = t_0$ can be written from Eq.\eqref{eq.1} in terms of node degrees as,
\begin{equation}\label{eq.2}
\begin{array}{l}
\dot{\bf x}_i(t_{0})=F_1({\bf x}_s)+\sum\limits_{\beta=1}^{M}\epsilon_\beta d_i^{[1,\beta]}(t) G_{\beta}^{[1]}({\bf x}_s,{\bf x}_s)  \\ ~~~~~~~~~~~~~+ \lambda e_{i}^{[1]} H_1({\bf x}_s,{\bf y}_s), \\
\dot{\bf y}_i(t_{0})=F_2({\bf y}_s)+\sum\limits_{\beta=1}^{M}\epsilon_\beta d_i^{[2,\beta]}(t) G_{\beta}^{[2]}({\bf y}_s,{\bf y}_s)  \\ ~~~~~~~~~~~~~+ \lambda e_{i}^{[2]}H_2({\bf y}_s,{\bf x}_s).
\end{array}
\end{equation}
To sustain intralayer synchronization solution, both the layers should advance on the same time evolution. Therefore, the velocities of any two distinct nodes in layer-1 should be identical, i.e.,
\begin{equation*}
\begin{array}{l}
\dot{\mathbf{x}}_i(t_{0}) = \dot{\mathbf{x}}_k(t_{0}), \;\; {\mbox{for any}~i\ne k,~\mbox{and}}~i,k=1,2,\dots,N.	
\end{array}
\end{equation*}
Since the intralayer and interlayer coupling functions are arbitrary, this yields for layer-1, 
\begin{equation*}
\begin{array}{l}
d_{i}^{[1,\beta]}(t)=d_{k}^{[1,\beta]}(t)$ \; \mbox{and} \; $e_{i}^{[1]}=e_{k}^{[1]}.
\end{array}
\end{equation*}
Similarly for layer-2, 
\begin{equation*}
\begin{array}{l}
d_{i}^{[2,\beta]}(t)=d_{k}^{[2,\beta]}(t)$ \; \mbox{and} \; $e_{i}^{[2]}=e_{k}^{[2]}.	
\end{array}
\end{equation*}
{\it Hence for the synchronization solution to be invariant the node degree of each dynamical unit for tier-$\beta $ in individual layers are identical and also the interlayer degree of all the nodes is equal for each layer.}

\par We further consider that the in-degree of nodes associated with tier-$\beta$ is constant in time, i.e.,
\begin{equation}{\label{eq.3}}
\begin{array}{l}
\sum\limits_{j=1}^{N}{\mathscr{A}}_{ij}^{[l,\beta]}(t)=d^{[\beta]}, l=1,2 \; \;\mbox{and} \;\; \beta=1,2,...,M.
\end{array}
\end{equation}
	
\par Hence the synchronous solutions $(\mathbf{x}_s,\mathbf{y}_s)$ evolves according to the following equations,
\begin{equation}\label{eq.4}
\begin{array}{lcl}
\dot{\bf x}_s = F_1({\bf x}_s)+\sum\limits_{\beta=1}^{M}\epsilon_\beta d^{[\beta]} G_{\beta}^{[1]}({\bf x}_s,{\bf x}_s) + \lambda e^{[1]}H_1({\bf x}_s,{\bf y}_s),\\
\dot{\bf y}_s = F_2({\bf y}_s)+\sum\limits_{\beta=1}^{M}\epsilon_\beta d^{[\beta]} G_{\beta}^{[2]}({\bf y}_s,{\bf y}_s) + \lambda e^{[2]}H_2({\bf y}_s,{\bf x}_s).
\end{array}
\end{equation}

\subsection{Stability Analysis}\label{Stability}
\par The intralayer synchronization occurs when the synchronous manifold $\mathcal{S}$ is stable under small perturbation in the transverse subspace. We consider small perturbation of the $i^{\mbox{th}}$ node around synchronous state, i.e.,
\begin{equation*}
\begin{array}{l}
\delta\mathbf{x}_i = \mathbf{x}_{i} - \mathbf{x}_{s} \;\mbox{and}\; \delta\mathbf{y}_i = \mathbf{y}_{i} - \mathbf{y}_{s},	
\end{array}
\end{equation*}
and execute linear stability analysis of Eq.\eqref{eq.1}. Therefore, the linearized equation can be expressed in terms of stack vectors as,	
\begin{equation}\label{eq.5}
\begin{array}{l}
\delta\dot{\mathbf{x}}=I_{N}{\otimes}JF_1(\mathbf{x}_s)\delta\mathbf{x}+\sum\limits_{\beta=1}^{M}\epsilon_{\beta} d^{[\beta]}[I_N{\otimes}JG_{\beta}^{[1]}(\mathbf{x}_s,\mathbf{x}_s)+\\~ I_N{\otimes}DG_{\beta}^{[1]}(\mathbf{x}_s,\mathbf{x}_s)]\delta\mathbf{x} -\sum\limits_{\beta=1}^{M}\epsilon_{\beta}\mathscr{L}^{[1,\beta]}(t){\otimes}DG_{\beta}^{[1]}({\bf x}_s,{\bf x}_s)\delta\mathbf{x}\\~~~~ +\lambda[e^{[1]}I_{N}{\otimes}JH_1(\mathbf{x}_s,\mathbf{y}_s)\delta\mathbf{x}+\mathscr{B}^{[1]}{\otimes}DH_1(\mathbf{x}_s,\mathbf{y}_s)\delta\mathbf{y}], \\\\

\delta\dot{\mathbf{y}}=I_{N}{\otimes}JF_2(\mathbf{y}_s)\delta\mathbf{y}+\sum\limits_{\beta=1}^{M}\epsilon_{\beta} d^{[\beta]}[I_N{\otimes}JG_{\beta}^{[2]}(\mathbf{y}_s,\mathbf{y}_s)+\\~ I_N{\otimes}DG_{\beta}^{[2]}(\mathbf{y}_s,\mathbf{y}_s)]\delta\mathbf{y} -\sum\limits_{\beta=1}^{M}\epsilon_{\beta}\mathscr{L}^{[2,\beta]}(t){\otimes}DG_{\beta}^{[1]}({\bf y}_s,{\bf y}_s)\delta\mathbf{y} \\~~~~ +\lambda[e^{[2]}I_{N}{\otimes}JH_2(\mathbf{y}_s,\mathbf{x}_s)\delta\mathbf{y}+\mathscr{B}^{[2]}{\otimes}DH_2(\mathbf{y}_s,\mathbf{x}_s)\delta\mathbf{x}],	
\end{array}
\end{equation} 
where, 
\begin{equation*}
\begin{array}{l}
\delta\mathbf{x}(t)=[\delta\mathbf{x}_{1}(t)^{tr}, \delta\mathbf{x}_{2}(t)^{tr},\dots,\delta\mathbf{x}_{N}(t)^{tr}]^{tr}, \; \mbox{and} \\\\ \delta\mathbf{y}(t)=[\delta\mathbf{y}_{1}(t)^{tr}, \delta\mathbf{y}_{2}(t)^{tr}, ..., \delta\mathbf{y}_{N}(t)^{tr}]^{tr}.	
\end{array}
\end{equation*}
$J$ and $D$ are the Jacobian derivative operators corresponding to the first and second variables, respectively. 

\par These linearized set of variational equations contains two components: one attributing for the motion along intralayer synchronization manifold, called parallel modes, and other describing the modes transverse to the manifold, called transverse modes. The necessary condition for synchronization state to be stable requires all the transverse modes must converge to zero in time. To perform the linear stability analysis one then needs to decouple the variational equation into parallel and transverse mode and examine whether the latter ones die out or not.  To accomplish stable synchronization state for static network, the classical MSF approach appropriately decouples the high-dimensional variational equation to perturbation mode independent low-dimensional equations through a suitable coordinate transformation which completely diagonalizes the coupling matrices. But for the temporal network scenario, as the Laplacian matrices changes with time, it accounts for many non-commuting matrices. Since, non-commuting matrices are not simultaneously diagonalizable, we can not directly apply the classical MSF formalism to these circumstances. For this reason, we review two different well-known techniques for stability analysis of our temporal multilayer hypernetwork. One is based on the fast switching stability technique, for which time-averaged networks are considered to bring down the problem of stability analysis in MSF form and the other one follows simultaneous block diagonal (SBD) approach, which simultaneously decouples an arbitrary set of symmetry matrices into finest block diagonalized form and as a result decouples the variational equation into optimal form without considering further conditions like time-averaged network formation.

\subsubsection{Fast switching approach} \label{fast_switching} 
\par Fast switching \cite{stilwell2006sufficient,belykh2004blinking,belykh2004connection} suggests that the change of network structure with time is much faster as compared to the evolution of oscillators coupled to each other. To perform the stability analysis we convert the variational equation \eqref{eq.5} to a suitable time-averaged form where the intralayer topologies corresponding to each tier are converted into time-averaged structures through the transformation of time-varying Laplacians into static time-averaged ones. In this regard, we recall the subsequent Lemma proposed by Stillwell et al. \cite{stilwell2006sufficient} regarding fast switching stability convention. 
\begin{lemma}\label{lemma_1}
If there exits a time average matrix $\bar{M}$ of the matrix valued function $M(t)$ such that
\begin{equation*}
\begin{array}{l}
\bar{M}=\frac{1}{T} \int_{t}^{t+T}{M(z)} dz,~\forall~t \in \mathbb{R}^{+},
\end{array}
\end{equation*}
and for some constant T, then for fairly fast switching the system,
\begin{equation}{\label{eq.6}}
\begin{array}{l}
\dot{\mathbf{w}}(t)=[A(t)+M(t)] \mathbf{w}(t),\hspace{10pt}\mathbf{w}(t_0)=\mathbf{w}_0,\hspace{10pt}t\ge t_0,
\end{array}
\end{equation}
will be uniformly asymptotically stable whenever the time average system,
\begin{equation}{\label{eq.7}}
\begin{array}{l}
\dot{\bar{\mathbf{w}}}(t)=[A(t)+\bar{M}] \bar{\mathbf{w}}(t), \hspace{10pt}\bar{\mathbf{x}}(t_0)=\bar{\mathbf{x}}_0, \hspace{10pt}t \ge t_0,
\end{array}
\end{equation}
is also uniformly asymptotically stable.
\end{lemma}
\par Using the above convention, we establish that the time-varying multilayer hypernetwork \eqref{eq.1} obtains stable intralayer synchronization state whenever the corresponding time-averaged system shows asymptotically stable {intralayer synchronization} state (The proof is illustrated in Appendix \ref{time_average}). As the network structures of tier $\beta$ for both the layers are equivalent, their corresponding time-averaged adjacency and Laplacian matrices are identical. Then for an arbitrary real constant T, 
\begin{equation*}
\begin{array}{l}
\frac{1}{T} \int_{t}^{t+T} \mathscr{A}^{[l,\beta]}(z) dz = \bar{\mathscr{A}^{[\beta]}}, \;\; \mbox{and} \\
\frac{1}{T} \int_{t}^{t+T} \mathscr{L}^{[l,\beta]}(z) dz = \bar{\mathscr{L}^{[\beta]}}, \;l=1,2\; \mbox{and}\; \beta=1,2,...,P.
\end{array}
\end{equation*}
Taking the time-average intralayer network topologies, the time-averaged multilayer hypernetwork can be expressed in terms of the following sets of equations as:	
\begin{equation}\label{eq.8}
\begin{array}{l}
\dot{{ \bar{\mathbf{x}}}}_i = F_1({\bf {\bar{x}}}_i)+\sum\limits_{\beta=1}^{M}\epsilon_\beta\sum\limits_{j=1}^{N}{\bar{\mathscr{A}}_{ij}^{[\beta]}} G_{\beta}^{[1]}({\bf \bar{x}}_i,{\bf \bar{x}}_j)  \\ ~~~~~~~~~~~~~+ \lambda \sum\limits_{j=1}^{N}{\mathscr{B}_{ij}^{[1]}}H_1({\bf \bar{x}}_i,{\bf \bar{y}}_j),\\
\dot{ \bar{\mathbf{y}}}_i = F_2({\bf \bar{y}}_i)+\sum\limits_{\beta=1}^{M}\epsilon_\beta\sum\limits_{j=1}^{N}{\bar{\mathscr{A}}_{ij}^{[\beta]}} G_{\beta}^{[2]}({\bf \bar{y}}_i,{\bf \bar{y}}_j)  \\ ~~~~~~~~~~~~~+ \lambda \sum\limits_{j=1}^{N}{\mathscr{B}_{ij}^{[2]}}H_2({\bf \bar{y}}_i,{\bf \bar{x}}_j),
\end{array}
\end{equation}
where $\mathbf{\bar{x}}_i(\mathbf{\bar{y}}_i)$ represents the states of the $i$-th node in layer-1 (layer-2) for time-averaged network. 

\par As, the stability of intralayer synchronization for time-average multilayer hypernetwork \eqref{eq.8} guarantees the stability of intralayer synchronization in time-varying network \eqref{eq.1}, the extinction of transverse modes corresponding to time-averages system will then necessary implies the achievement of stable coherent state. Through a series of theoretical steps detailed in Appendix \ref{time_average}, we derive the dynamics of transverse modes for time-average system as following:
\begin{equation}\label{eq.9}
\begin{array}{l}
\dot{\eta}_T^{(\mathbf{\bar{x}})}=I_{N-1}{\otimes}JF_1(\mathbf{x}_s)\eta_T^{(\mathbf{\bar{x}})} +\sum\limits_{\beta=1}^{M}\epsilon_{\beta} d^{[\beta]}[I_{N-1}{\otimes}JG_{\beta}^{[1]}(\mathbf{x}_s,\mathbf{x}_s) \\~+I_{N-1}{\otimes}DG_{\beta}^{[1]}(\mathbf{x}_s,\mathbf{x}_s)]\eta_T^{(\mathbf{\bar{x}})}   -\sum\limits_{\beta=1}^{M}\epsilon_{\beta}\bar{W}_2^{[\beta]}{\otimes}DG_{\beta}^{[1]}({\bf x}_s,{\bf x}_s)\eta_T^{(\mathbf{\bar{x}})}\\~~ +\lambda[e^{[1]}I_{N-1}{\otimes}JH_1(\mathbf{x}_s,\mathbf{y}_s)\eta_T^{(\mathbf{\bar{x}})}+U_3^{[1]}{\otimes}DH_1(\mathbf{x}_s,\mathbf{y}_s)\eta_T^{(\mathbf{\bar{y}})}],  \\\\

\dot{\eta}_T^{(\mathbf{\bar{y}})}=I_{N-1}{\otimes}JF_2(\mathbf{y}_s)\eta_T^{(\mathbf{\bar{y}})}+\sum\limits_{\beta=1}^{M}\epsilon_{\beta} d^{[\beta]}[I_{N-1}{\otimes}JG_{\beta}^{[2]}(\mathbf{y}_s,\mathbf{y}_s)\\~~~~ +I_{N-1}{\otimes}DG_{\beta}^{[2]}(\mathbf{y}_s,\mathbf{y}_s)]\eta_T^{(\mathbf{\bar{y}})}  \\~~~~~~ -\sum\limits_{\beta=1}^{M}\epsilon_{\beta}\bar{W}_2^{[\beta]}{\otimes}DG_{\beta}^{[2]}({\bf y}_s,{\bf y}_s)\eta_T^{(\mathbf{\bar{y}})} \\~~~~+\lambda[e^{[2]}I_{N-1}{\otimes}JH_2(\mathbf{y}_s,\mathbf{x}_s)\eta_T^{(\mathbf{\bar{y}})}+U_3^{[2]}{\otimes}DH_2(\mathbf{y}_s,\mathbf{x}_s)\eta_T^{(\mathbf{\bar{x}})}],
\end{array}
\end{equation}
where ${\eta}_T^{(\mathbf{\bar{x}})}$ and ${\eta}_T^{(\mathbf{\bar{y}})}$ represents the states of transverse modes of time-average system.

\par Therefore, the problem of stability of intralayer synchronization state for the time-varying multilayer network is then reduced to solving the coupled transverse linear equations \eqref{eq.9} for calculation of maximum Lyapunov exponent (MLE). Stability of coherent state needs MLE associated with the transverse modes to be negative, as a necessary condition. Given node dynamics and coupling functions, MLE is mainly function of coupling strengths, time-averaged Laplacians associated with each tier and interlayer adjacencies, i.e., MLE ($\epsilon_{1}, \epsilon_{2},\dots,\epsilon_{M},\lambda,\bar{\mathscr{L}}^{[1]}, \bar{\mathscr{L}}^{[2]},\dots,\bar{\mathscr{L}}^{[M]},\mathscr{B}^{[1]},\mathscr{B}^{[2]}$). It is notable that, in comparison with the classical MSF scheme, also in time-varying multilayer hypernetwork we are able to separate the motion parallel and transverse to synchronization manifold through fast switching approach. However, more intricacy in network structure in latter circumstance gives a bunch of coupled linear differential equation to analyze the stability, in spite of independent, uncoupled equations as in the case of classical MSF.  Particularly, there remain some terms in the transverse error equation \eqref{eq.9} which are generally not transformable into block diagonal forms and as a result the transverse error dynamics becomes $2d(N-1)$-dimensional coupled equation. Furthermore, there are suitable occasions where the coupled transverse modes can be optimally separated and the $2d(N-1)$-dimensional coupled equation reduce into $(N-1)$, $2d$-dimensional linear differential equations. 

\par The relevant instances are as follows:\\
(i) The first instance is when each time average Laplacians $\bar{\mathscr{L}}^{[\beta]}$, interlayer adjacency matrices $\mathscr{B}^{[l]}$, $l=1,2 \;\;\mbox{and}\;\; \beta= 1, 2,\dots,M$ are symmetric, and among them one Laplacian commutes with all the other time average Laplacians as well as with the interlayer adjacencies (The detailed proof is in Appendix \ref{commutative}). In this case the transverse variational equation reduces to $2d$-dimensional $N-1$ systems as,
\begin{equation}\label{eq.10}
\begin{array}{l}
\dot{\eta}_{T_i}^{(\mathbf{\bar{x}})}=JF_1(\mathbf{x}_s)\eta_{T_i}^{(\mathbf{\bar{x}})}+\sum\limits_{\beta=1}^{M}\epsilon_{\beta} d^{[\beta]}[JG_{\beta}^{[1]}(\mathbf{x}_s,\mathbf{x}_s) \\~~~~ +DG_{\beta}^{[1]}(\mathbf{x}_s,\mathbf{x}_s)]\eta_{T_i}^{(\mathbf{\bar{x}})} -\sum\limits_{\beta=1}^{M}\epsilon_{\beta}\bar{\gamma}_{i}^{[\beta]}DG_{\beta}^{[1]}({\bf x}_s,{\bf x}_s)\eta_{T_i}^{(\mathbf{\bar{x}})} \\~~~~ +\lambda[e^{[1]}JH_1(\mathbf{x}_s,\mathbf{y}_s)\eta_{T_i}^{(\mathbf{\bar{x}})}+\varGamma_{i}^{[1]}DH_1(\mathbf{x}_s,\mathbf{y}_s)\eta_{T_i}^{(\mathbf{\bar{y}})}], \\

\dot{\eta}_{T_i}^{(\mathbf{\bar{y}})}=JF_2(\mathbf{y}_s)\eta_{T_i}^{(\mathbf{\bar{y}})}+\sum\limits_{\beta=1}^{M}\epsilon_{\beta} d^{[\beta]}[JG_{\beta}^{[2]}(\mathbf{y}_s,\mathbf{y}_s) \\~~~~ +DG_{\beta}^{[2]}(\mathbf{y}_s,\mathbf{y}_s)]\eta_{T_i}^{(\mathbf{\bar{y}})}  -\sum\limits_{\beta=1}^{M}\epsilon_{\beta}\bar{\gamma}_{i}^{[\beta]}DG_{\beta}^{[2]}({\bf y}_s,{\bf y}_s)\eta_{T_i}^{(\mathbf{\bar{y}})} \\~~~~ +\lambda[e^{[2]}JH_2(\mathbf{y}_s,\mathbf{x}_s)\eta_{T_i}^{(\mathbf{\bar{y}})}+\varGamma_{i}^{[2]}DH_2(\mathbf{y}_s,\mathbf{x}_s)\eta_{T_i}^{(\mathbf{\bar{x}})}], 
\end{array}
\end{equation}
where $\bar{\gamma}_{i}^{[\beta]}$ and $\varGamma_{i}^{[l]}$ for $i=2,3,\dots,N$ are the non-zero eigenvalues of time-averaged Laplacian matrices corresponding to tier-$\beta$ and eigenvalues of interlayer adjacency matrices.  

\par As for example, if we consider connection topology of one tier of time-varying network as random network with edge rewiring probability $p_{rand}$ and any arbitrary undirected network structure for other tiers with bidirectional interlayer connections, then we can get one such instance where transverse dynamics decouples into independent modes.\\
(ii) Another instance for optimal separation of transverse modes occurs when the number of distinct connection topologies in both layer equals to one, i.e., M = 1 and the time average Laplacian matrices have all the eigenvalues equal except the smallest zero eigenvalue. Further the interlayer adjacencies  $\mathscr{B}^{[1]}$ and $\mathscr{B}^{[2]}$ are identical (see Appendix \ref{one_tier} for full derivation).

\subsubsection{Simultaneous block diagonalization approach} \label{sbd}
\par To separate the parallel and transverse error components the fast switching approach considers the time-averaged network structures. The commutative property of time-average Laplacians and identical eigenvalues are also considered to further reduce the transverse error system into lower dimensional form. To overcome these limitations, here we use the SBD \cite{maehara2011algorithm,sbd_sorrentino,zhang2020symmetry} approach on the set of coupling matrices of time-varying multilayer hypernetwork \eqref{eq.1}.

\par The {SBD} process can be conceptualized as follows: for a prescribed {finite} set of symmetric matrices $\{B^{1}, B^{2},\dots,B^{q}\}$, there exists an orthogonal matrix $P$ such that the matrices $P^{tr}B^{m}P$ possess a common block-diagonal structure for $m=1,2,\dots,q$, leading to an optimal separation of perturbation modes. This common block structure is identified by the largest number of blocks and is unique up to block permutations. In particular, this scheme guarantees the separation of perturbation modes along the synchronization manifold and transverse to the manifold. Further it decouples transverse system into low-dimensional systems through transforming set of matrices to optimal block diagonal forms.   

\par To proceed, we rewrite the variation equation \eqref{eq.5} by introducing $\delta \mathbf{ X}(t) =\begin{bmatrix} \delta \mathbf{x}(t) & \delta \mathbf{y}(t) \end{bmatrix}^{tr}$ as,
\begin{equation} \label{eq.11}
\begin{array}{lll}
{\delta\dot{\mathbf X}}(t)=\begin{bmatrix}
\delta\dot{\mathbf{x}}(t) \\ \delta\dot{\mathbf{y}}(t) 
\end{bmatrix} = \big[ D_{1}
+\sum\limits_{\beta=1}^{M}\epsilon_{\beta} d^{[\beta]} D_{2} + \sum\limits_{\beta=1}^{M}\epsilon_{\beta} D_{3}\\\hspace{82pt}+\sum\limits_{\beta=1}^{M}\epsilon_{\beta} D_{4}  + \lambda D_{5} +\lambda D_{6} \big] \delta{\mathbf X},
\end{array}
\end{equation}
where \begin{equation*}
		\begin{array}{l}
		D_{1}=\begin{bmatrix}
			I_{N}\otimes JF_{1} & 0 \\ 0  & I_{N} \otimes JF_{2}
		\end{bmatrix},
	  \\\\D_{2}= \begin{bmatrix}
		I_{N}\otimes JG_{\beta}^{[1]}+I_{N}\otimes DG_{\beta}^{[1]} & 0 \\
		0 & I_{N}\otimes JG_{\beta}^{[2]}+I_{N}\otimes DG_{\beta}^{[2]}
	\end{bmatrix}, \\\\D_{3}= \begin{bmatrix}
		\mathscr{L}^{[1,\beta]}(t) \otimes DG_{\beta}^{[1]} & 0 \\ 
		0 & 0
	\end{bmatrix}, \\\\D_{4}= \begin{bmatrix}
		0 & 0 \\
		0 & \mathscr{L}^{[2,\beta]}(t) \otimes DG_{\beta}^{[2]}
	\end{bmatrix}, \\\\D_{5}= \begin{bmatrix}
		e^{[1]}I_{N}\otimes JH_{1} & 0 \\ 
		0 & e^{[2]}I_{N}\otimes JH_{2}
	\end{bmatrix}, \\\\ \;\; \mbox{and}\;\; D_{6}= \begin{bmatrix}
		0 & \mathscr{B}^{[1]}\otimes DH_{1}\\
		\mathscr{B}^{[2]}\otimes DH_{2} & 0
	\end{bmatrix}.
    \end{array}
    \end{equation*}
	
	Now, let 
	 \begin{equation*}
	 	\begin{array}{l}
	 		U^{[1,\beta]}(t)= \begin{bmatrix}
	 			\mathscr{L}^{[1,\beta]}(t) & 0 \\
	 			0 & 0 
	 		\end{bmatrix} ,  \\\\ U^{[2,\beta]}(t)= \begin{bmatrix}
	 			0 & 0 \\
	 			0 & \mathscr{L}^{[2,\beta]}(t)
	 		\end{bmatrix}\;\; \mbox{and} \;\;\mathscr{B}= \begin{bmatrix}
	 			0 & \mathscr{B}^{[1]} \\
	 			\mathscr{B}^{[2]}  & 0 
	 		\end{bmatrix}.
	 	\end{array}
	 \end{equation*}  Also, we consider that the interlayer coupling functions hold $JH_{1}(x_0,yo)= JH_{2}(y_0,x_o)$ and $DH_{1}(x_0,y_o)= DH_{2}(y_0,x_o)$.
	
	
	Therefore, the variational equation \eqref{eq.11} can be rewritten as, 
	
	\begin{equation} \label{eq.12}
		\begin{array}{l}
			{\delta\dot{\mathbf X}} = \big[ D_{1}
			+ \sum\limits_{\beta=1}^{M}\epsilon_{\beta} d^{[\beta]} D_{2} + \sum\limits_{\beta=1}^{M}\epsilon_{\beta} U^{[1,\beta]}(t) \otimes DG_{\beta}^{[1]}  \\ ~~~~~~~~+ 
			\sum\limits_{\beta=1}^{M}\epsilon_{\beta} U^{[2,\beta]}(t) \otimes DG_{\beta}^{[2]}  + \lambda D_{5} +\lambda \mathscr{B}\otimes DH_1 \big] \delta{\mathbf X}.
		\end{array}
	\end{equation}
All the terms in this linearized equation are block diagonalized except the terms $U^{[1,\beta]}(t) \otimes DG_{\beta}^{[1]}$, $U^{[2,\beta]}(t) \otimes DG_{\beta}^{[2]}$ and $\mathscr{B}\otimes DH_1$. Now time-varying Laplacian matrices $\mathscr{L}^{[1,\beta]}(t)$, $\mathscr{L}^{[2,\beta]}(t)$ and the interlayer adjacency matrices  are all symmetric matrices. Hence the matrices $U^{[1,\beta]}(t)$, $U^{[2,\beta]}(t)$ and $\mathscr{B}$ are also symmetric. So, we apply the process of SBD on the set of symmetric matrices \{$U^{[1,\beta]}(t),U^{[2,\beta]}(t),\mathscr{B}$\} to obtain a orthogonal matrix $P$ that transforms these matrices to finest block diagonalization form. {Usually the SBD technique works better for a finite collection of symmetric matrices, so we adopt one extra assumption about the time varying intralayer connection topologies. We assume that the collection of time-stamped intralayer adjacency matrices $\{\mathscr{A}^{[l,\beta]}(t):~l=1,2;~\beta=1,2,\dots,M;~\mbox{and}~t\in\mathbb{R}^+\}$ is a finite set, which in turn gives finite number of symmetric matrices $U^{[1,\beta]}(t),U^{[2,\beta]}(t)$}. To obtain the matrix $P$, we follow the algorithm illustrated in \cite{zhang2020symmetry}. Thereafter, we project the state variable $\delta \mathbf{X}$ onto the basis of vectors formed by the columns of $P$ by defining new variable, 
\begin{equation*}
\begin{array}{l}
\eta(t)=(P{\otimes}I_{2d})^{T}\delta\mathbf{ X}(t).
\end{array}
\end{equation*} Using this co-ordinate transformation the linearized equation \eqref{eq.12} becomes,
\begin{equation} \label{eq.13}
\begin{array}{l}
\dot{\eta}(t)=\big[ D_{1}+\sum\limits_{\beta=1}^{M}\epsilon_{\beta} d^{[\beta]} D_{2}+\sum\limits_{\beta=1}^{M}\epsilon_{\beta} \tilde{U}^{[1,\beta]}(t) \otimes DG_{\beta}^{[1]}+  \\ ~~~~~~~~ \sum\limits_{\beta=1}^{M}\epsilon_{\beta} \tilde{U}^{[2,\beta]}(t) \otimes DG_{\beta}^{[2]}+ \lambda D_{5}+\lambda \tilde{\mathscr{B}}\otimes DH_1 \big] \eta(t),
\end{array}
\end{equation} 	
where
\begin{equation*}
\begin{array}{lll}
\tilde{U}^{[1,\beta]}(t) &= P^{T} U^{[1,\beta]}(t) P, \\ \tilde{U}^{[2,\beta]}(t) &= P^{T} U^{[2,\beta]}(t) P, \\ \tilde{\mathscr{B}} &= P^{T} \mathscr{B} P,
\end{array}
\end{equation*}
are the finest block diagonalized forms of $U^{[1,\beta]}(t)$,$U^{[2,\beta]}(t)$ \& $\mathscr{B}$. These block diagonalized matrices have two $1\times1$ blocks which are associated with the parallel modes of synchronization solutions corresponding to two layers and the other blocks are associated with the transverse modes of synchronization manifold. Thus, we can perfectly decouple the perturbation modes parallel and transverse to the synchronization manifold without implementing time-average network structure. Apart from this, the finest block-diagonal forms of the matrices also explains that the transverse modes are not fully coupled. The high-dimensional coupled transverse system reduces to many lower-dimensional linear systems by means of sizes of blocks in the transformed block-diagonal matrices. Hence, the further reduction of transverse error systems is possible through this approach even without considering any limitations like fast switching mechanism. The stability condition of the synchronization manifold is then reduced to solving the linear equation \eqref{eq.13} for calculation of maximum Lyapunov exponent transverse to synchronous manifold. The necessary condition requires the maximum Lyapunov exponent to be negative. In Sec. \ref{sbd_numerical}, we have elaborated the whole process with a suitable example.
	
	\section{Numerical Illustrations}\label{numerical}
	
		
	In order to illustrate our theoretical findings, here we consider three cases of different isolate node dynamics and nonlinear coupling functions. Specifically, we address two ideal chaotic systems, the Lorenz \cite{strogatz2018nonlinear} and R\"{o}ssler \cite{rossler1976equation} systems, and as a real-world instance of neuronal evolution, the Sherman model \cite{sherman1994anti,jalil2012spikes}. For numerical simulations, we consider that each layer of the multilayer network consists of $N$ number of nodes interacting through two structurally different tiers at every instance of time. To keep things simple, we take the interlayer connections as one to one between the nodes of adjacent layers and dynamics of isolated nodes in both layers are identical. In order to investigate the intralayer synchronous state, we define the synchronization error as,
	\begin{equation} \label{eq.14}
		E_{intra}=\lim\limits_{T\to\infty}\dfrac{1}{T}\int_{t_{trans}}^{t_{trans}+T}E_{intra}(t)~dt,
	\end{equation}
     where,~~$E_{intra}(t)=\sum\limits_{k=2}^{N}\dfrac{\|{\bf x}_k(t)-{\bf x}_1(t)\|+\|{\bf y}_k(t)-{\bf y}_1(t)\|}{2(N-1)}$.
	
\par Here $\|\cdot\|$ symbolizes the Euclidean norm, $t_{trans}$ determines the transient of the numerical simulation and $T$ is a sufficiently large positive number. Asymptotic stability of $E_{intra}$ will imply each layer is synchronized. For intralayer synchrony to emerge, we have used the threshold value of $E_{intra}$ as $10^{-5}$. Numerical simulations of the multilayer network are executed using  the $4^{th}$ order Runge-Kutta algorithm, with step of integration $\delta t$= $0.01$ (for chaotic oscillators) and $0.001$ (for neuron model), over a span of time $T=1000$ after an initial transient $t_{trans}=2000$. The initial condition of each dynamical node is chosen randomly from the phase-space of isolate node dynamics.  At each instant, we rewire all tiers of both layers separately with probability $f \delta t$, where $f$ is the frequency of rewiring process. All the numerical results are obtained by taking average over 10 network realizations and initial conditions. {At first, we will present the numerically obtained results based on the fast-switching stability criterion.}

\subsection{Numerical illustration of the fast switching approach}
\par In the next three sub-sections, we investigate the synchronization state numerically by means of synchronization error $E_{intra}$ and concurrently validate our analytical findings corresponding to fast switching approach for the three structurally different networked systems with $N=200$ nodes and different nonlinear coupling functions in each occasion.

\subsubsection{Coupled Lorenz systems with sine coupling functions} \label{lorenz_numerical}
\par We first consider multilayer hypernetwork consisting of coupled Lorenz oscillators interacting with sine coupling functions. Then the corresponding  equation of motion reads as:
\begin{equation} \label{eq.15}
\begin{array}{lll}
	
	\mbox{Layer-1 :} \\
\dot{x}_{1i}=\sigma(y_{1i}-x_{1i})+\epsilon_1\sum\limits_{j=1}^{N}\mathscr{A}^{[1,1]}_{ij}(t)\sin(x_{1j}-x_{1i}) \\~~~~~~~~~~+\lambda \sin(x_{2i}-x_{1i}),\\
\dot{y}_{1i}=x_{1i}(\rho-z_{1i})-y_{1i}+\epsilon_2\sum\limits_{j=1}^{N}\mathscr{A}^{[1,2]}_{ij}(t)\sin(y_{1j}-y_{1i}),\\
\dot{z}_{1i}=x_{1i}y_{1i}-\alpha z_{1i},\\\\

 \mbox{Layer-2 :} \\
\dot{x}_{2i}=\sigma(y_{2i}-x_{2i})+\epsilon_1\sum\limits_{j=1}^{N}\mathscr{A}^{[2,1]}_{ij}(t)\sin(x_{2j}-x_{2i})\\~~~~~~~~~~+\lambda \sin(x_{1i}-x_{2i}),\\
\dot{y}_{2i}=x_{2i}(\rho-z_{2i})-y_{2i}+\epsilon_2\sum\limits_{j=1}^{N}\mathscr{A}^{[2,2]}_{ij}(t)\sin(y_{2j}-y_{2i}),\\
\dot{z}_{2i}=x_{2i}y_{2i}-\alpha z_{2i},\hspace{50pt}i=1,2,\dots,N.
\end{array}
\end{equation}
where each isolate unit is in a chaotic state with system parameters $\sigma=10$, $\alpha=8/3$, and $\rho=28$. The intralayer connection topology for tier-1 is represented by Erd{\"o}s-R{\'e}nyi random network \cite{erdos2011evolution} with probability $p_{rand}$ and for tier-2 the intralayer connection topology is small-world network \cite{watts1998collective} with average node degree $2k_{sw}$ and link rewiring probability $p_{sw}$.

\begin{figure}[ht]
\centerline{\includegraphics[scale=0.43]{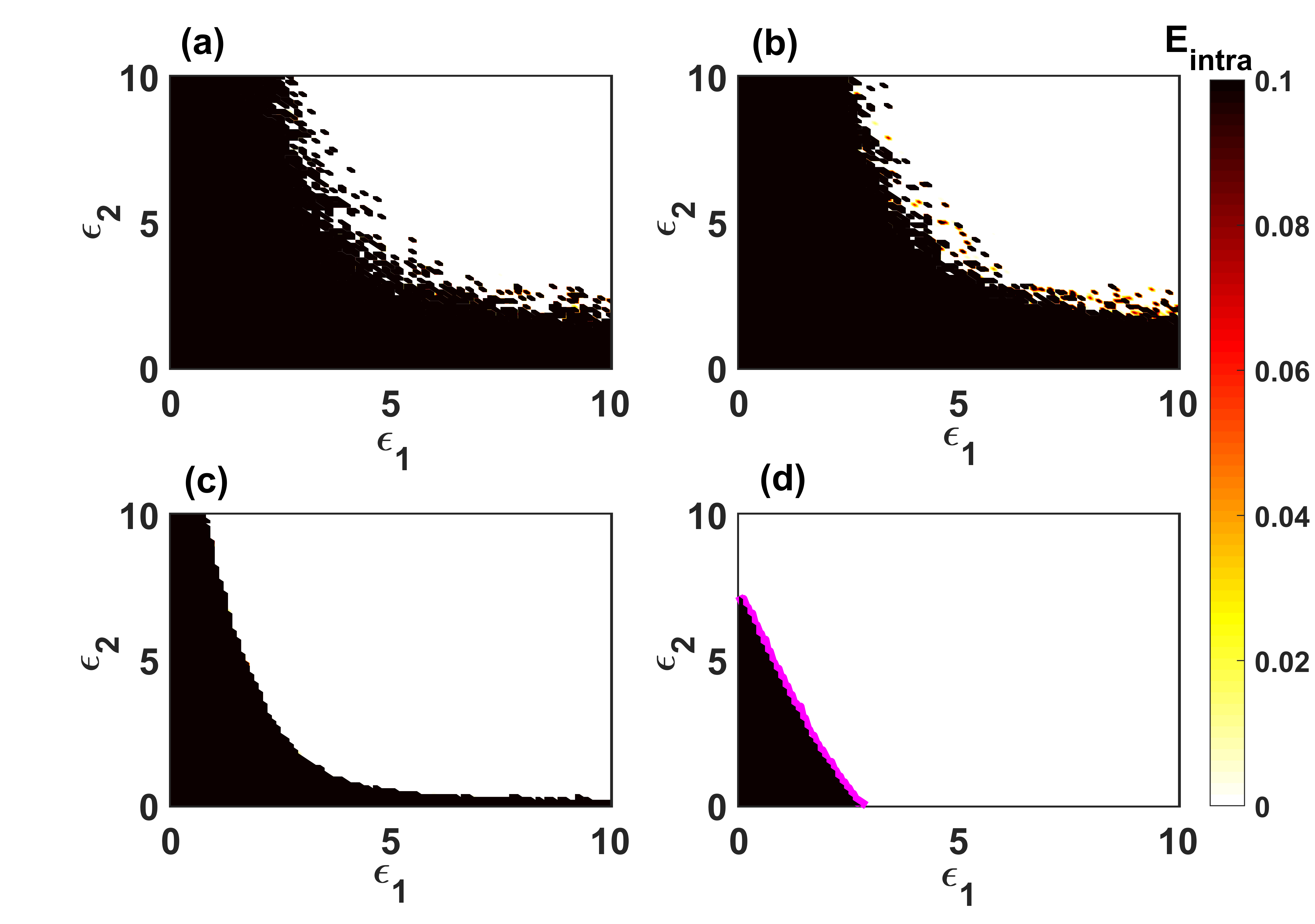}}
\caption{{\bf Coupled Lorenz oscillators:} Intralayer synchronization error in ($\epsilon_{1}$,$\epsilon_{2}$) parameter plane. For (a) $f=10^{-4}$, (b) $f=10^{-2}$, (c) $f=10^{0}$ and (d) $f=10^{2}$. Other parameter values are $d_{sw}=4$, $p_{sw}=0.1$, $p_{rand}=0.015$, $\lambda=0.1$. The solid {magenta} line in (d) is the theoretical prediction of intralayer synchronization threshold corresponding to MLE $=0$. White region represents stable intralayer synchronization state.}\label{fig2}
\end{figure}
For various rewiring frequencies, the intralayer synchronization error as a function of tier-1 coupling strength $\epsilon_{1}$ and tier-2 coupling strength $\epsilon_{2}$ is depicted in Fig. \ref{fig2}. The coherence and incoherence regions, portrayed in white and black, are plotted in Figs. $\ref{fig2}(a)$- $\ref{fig2}(d)$ for slow rewiring frequency $(f=10^{-4})$ to very fast rewiring frequency $(f=10^{2})$, respectively. For lower rewiring frequency (Fig. \ref{fig2}(a)), i.e, when the intralayer links are nearly steady with time, higher tier-2 coupling strength is needed to achieve synchrony and critical transition point of $\epsilon_2$ decrease as the value of $\epsilon_1$ increases. In Fig. \ref{fig2}(b) when rewiring frequency f is increased to $f=10^{-2}$, a small enhancement of synchronization region is observed in $(\epsilon_1,\epsilon_2)$ plane. For further increment of $f=10^0$ and $f=10^2$, a notable enhancement in coherent region is perceived in Figs. \ref{fig2}(c) and \ref{fig2}(d). 

\begin{figure}[ht]
\centerline{\includegraphics[scale=0.42]{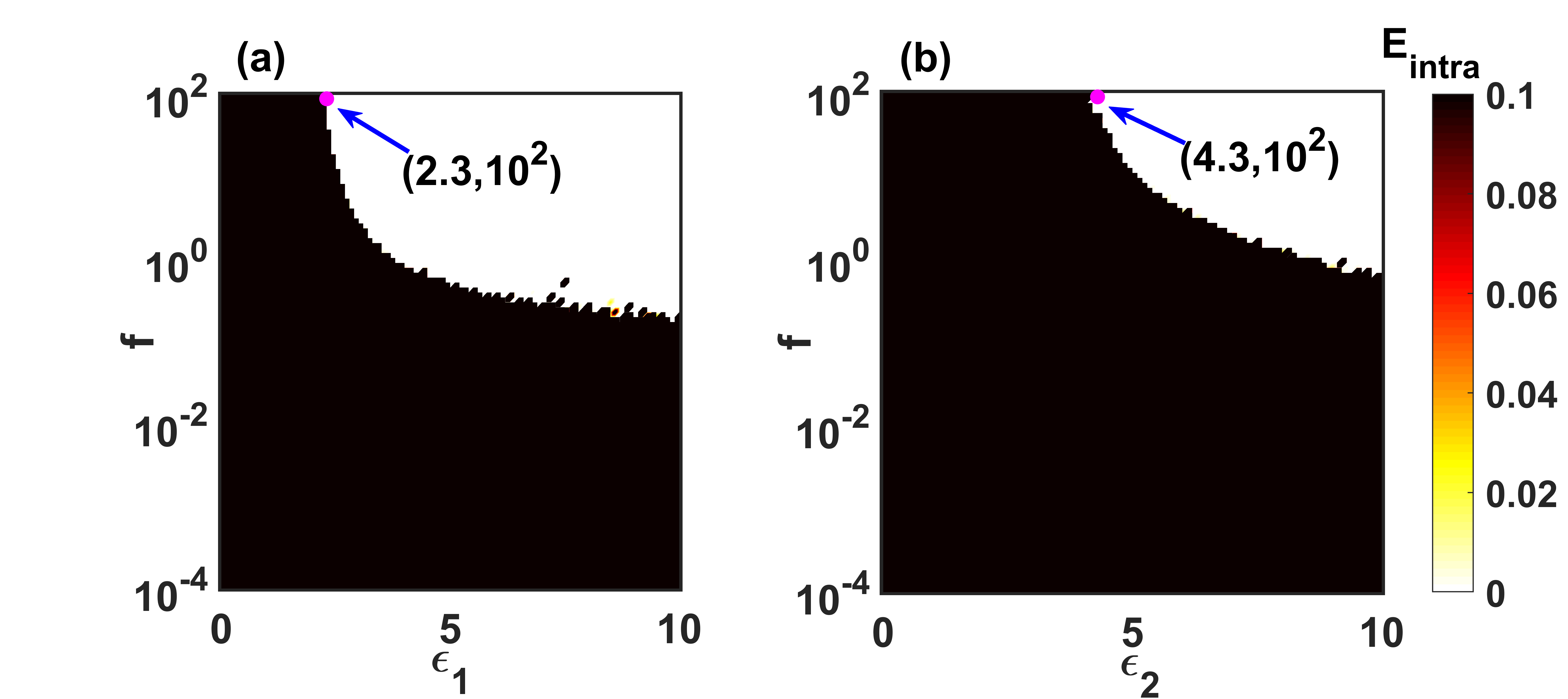}~~~~~}
\caption{{\bf Coupled Lorenz oscillators:} Intralayer synchronization error in (a) ($\epsilon_{1}$,f) and (b)  ($\epsilon_{2}$,f)  plane. For (a) $\epsilon_{2}=1.0$ and for (b)  $\epsilon_{1}=1.0$. Other parameter values are $d_{sw}=4$, $p_{sw}=0.1$, $p_{rand}=0.015$, $\lambda=0.1$. { The solid magenta circles in both the subfigures indicates analytically derived critical value of coupling strengths for sufficiently fast switching of intralayer network topologies. White region represents stable intralayer synchronization state. }}\label{fig3}
\end{figure}
\par Figure \ref{fig3} portrays variation of intralayer synchronization in the parameter plane $(\epsilon_1,f)$ and $(\epsilon_2,f)$. For $\epsilon_2=1.0$, the synchronized and desynchronized regions are plotted by varying $\epsilon_1$ and $f$ in Fig. \ref{fig3}(a). At small values of $\epsilon_1$ up to $2.5$, no synchronization occurs by varying rewiring frequency from slow to fast switching. Beyond $\epsilon_1=2.5$, higher rewiring frequency required to achieve synchrony and the critical point of $f$ reduces with increasing value of $\epsilon_1$. Similarly, Fig. \ref{fig3}(b) demonstrates the coherent and incoherent domains in the $(\epsilon_2,f)$ plane for fixed value of $\epsilon_1=1.0$.         
	
	\par Next we proceed through the stability of synchronized state based on the fast switching approach. If $\bar{\mathscr{L}}^{[1]}$ and $\bar{\mathscr{L}}^{[2]}$ are the time-average Laplacian matrices corresponding to tier-1 (random network) and tier-2 (small world network) for both the layers, then 
	\begin{equation*}
		\bar{\mathscr{L}}_{ij}^{[1]}= \begin{cases}
			-p_{rand}, & \mbox{for} \;\;i\neq j \\
			(N-1)p_{rand}, & \mbox{for} \;\;i = j
		\end{cases}
	\end{equation*} 
	and 
	
	\begin{equation*}
		\bar{\mathscr{L}}_{ij}^{[2]}= \begin{cases}
			-(1-p_{sw}), & \mbox{for} \;\; i-k_{sw}\leq j \leq i+k_{sw} \;\; \mbox{and} \;\; i\neq j \\
			2k_{sw}, & \mbox{for} \;\; i = j \\
			-\frac{2k_{sw}p_{sw}}{N-2k_{sw}-1}, & \mbox{otherwise}.
		\end{cases}
	\end{equation*} 
	
Clearly $\bar{\mathscr{L}}^{[1]}$ commutes with the other Laplacian $\bar{\mathscr{L}}^{[2]}$ and the interlayer adjacency matrix (in this case identity matrix). So, we can recast the transverse error equation in the reduced form as Eq.\eqref{eq.10}, which in this case becomes,
\begin{equation} \label{eq.16}
\begin{array}{l}
\delta\dot{x}_{1i}= \sigma(\delta y_{1i}-\delta x_{1i}) -\epsilon_{1}\bar{\gamma}_{i}^{[1]}\delta x_{1i}     \\~~~~~~~~~~+ \lambda \cos(x_2-x_1)(\delta x_{2i}-\delta x_{1i}),     \\
\delta\dot{y}_{1i}=(\rho-z_1)\delta x_{1i}-\delta y_{1i} -x_{1}\delta z_{1i}-\epsilon_2 \bar{\gamma}_{i}^{[2]} \delta y_{1i}, \\
\delta\dot{z}_{1i}=y_{1}\delta x_{1i} +x_{1}\delta y_{1i}-\beta \delta z_{1i}, \\\\
\delta\dot{x}_{2i}= \sigma(\delta y_{2i}-\delta x_{2i}) -\epsilon_{1} \bar{\gamma}_{i}^{[1]}\delta x_{2i}   \\~~~~~~~~~~+\lambda \cos(x_1-x_2)(\delta x_{1i}-\delta x_{2i}),     \\
\delta\dot{y}_{2i}=(\rho-z_2)\delta x_{2i}-\delta y_{2i} -x_{2}\delta z_{2i}-\epsilon_2 \bar{\gamma}_{i}^{[2]} \delta y_{2i}, \\
\delta\dot{z}_{2i}=y_{2}\delta x_{2i} +x_{2}\delta y_{2i}-\beta \delta z_{2i},   \hspace{10 pt} i=2,...,N,
\end{array}
\end{equation}
where $(x_1,y_1,z_1)$ and $(x_2,y_2,z_2)$ are the state variables of the synchronization manifold.

\par The necessary condition for synchronization requires maximum Lyapunov exponent obtained by simulating Eq.\eqref{eq.16}, to be negative. We verify the numerically obtained intralayer synchronization region in the $(\epsilon_1,\epsilon_2)$ plane by means of maximum transverse Lyapunov exponent for adequately fast switching. The critical curve for synchronization, characterized by MLE = 0 is drawn in solid {magenta} line in Fig. \ref{fig2}(d). In this parameter space, the regions above and below the threshold curve respectively depict coherent and incoherent state, which exactly agrees with the numerically obtained result on the basis of $E_{intra}$. Hence, for sufficiently fast switching the stability analysis of coherent state through the time-averaged network formation exactly signifies the coherence in time-varying network.

\subsubsection{Coupled R\"{o}ssler systems with nonlinear coupling functions}
\par To further elucidate that the time-varying interactions significantly contributes for the emergence of intralayer synchronization, we consider the multilayer hypernetwork \eqref{eq.1} with isolated node dynamics as the R\"{o}ssler oscillators, interacting with nonlinear coupling functions. Then the equation of motion of the multilayer network can be described as follows
	\begin{equation} \label{eq.17}
		\begin{array}{lll}
			
			\mbox{Layer-1 :} \\
			\dot{x}_{1i}=-y_{1i}-z_{1i}+\epsilon_1\sum\limits_{j=1}^{N}\mathscr{A}^{[1,1]}_{ij}(t)(\alpha-x_{1i})(x_{1j}-\beta)^2,\\
			\dot{y}_{1i}=x_{1i}+ay_{1i}+\epsilon_2\sum\limits_{j=1}^{N}\mathscr{A}^{[1,2]}_{ij}(t)(\alpha-y_{1i})(y_{1j}-\beta)^2 \\~~~~~~~~~~+\lambda(\alpha-y_{1i})(y_{2i}-\beta)^2,\\
			\dot{z}_{1i}=b+(x_{1i}-c)z_{1i},\\\\
			
			\mbox{Layer-2 :} \\
			\dot{x}_{2i}=-y_{2i}-z_{2i}+\epsilon_1\sum\limits_{j=1}^{N}\mathscr{A}^{[2,1]}_{ij}(t)(\alpha-x_{2i})(x_{2j}-\beta)^2,\\
			\dot{y}_{2i}=x_{2i}+ay_{2i}+\epsilon_2\sum\limits_{j=1}^{N}\mathscr{A}^{[2,2]}_{ij}(t)(\alpha-y_{2i})(y_{2j}-\beta)^2\\~~~~~~~~~~+\lambda(\alpha-y_{2i})(y_{1i}-\beta)^2,\\
			\dot{z}_{2i}=b+(x_{2i}-c)z_{2i},\hspace{20pt}i=1,2,\dots,N.
		\end{array}
	\end{equation}
	The system parameter values are chosen as $a=0.2$, $b=0.2$, $c=5.7$, and the two coupling parameter values chosen as $\alpha=0.37$, $\beta=-0.37$. The intralayer connection mechanisms for tier-1 and tier-2 are considered as random network with constant node degree $k_1=2$ and $k_2=3$ to guarantee the invariance condition of intralayer synchronization. The intralayer coupling functions are acting through $x$ and $y$ variable corresponding to tier-1 and tier-2, respectively, and interlayer coupling is through $y$ variable.

\begin{figure}[ht]
\centerline{\includegraphics[scale=0.425]{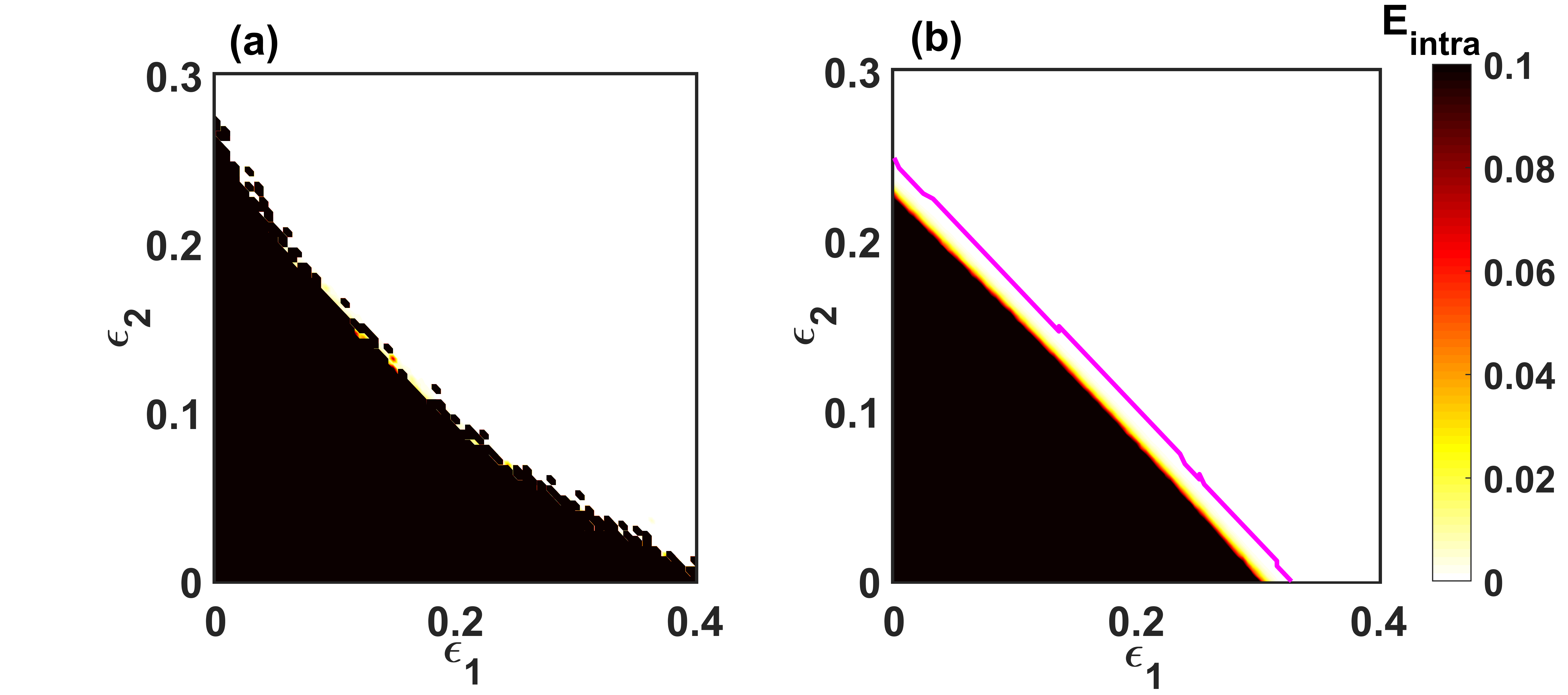}~~~~~~~~}
\caption{{\bf Coupled R\"{o}ssler oscillators:} Intralayer synchronization error in ($\epsilon_{1}$,$\epsilon_{2}$) parameter plane. For (a) $f=10^{-4}$, (b) $f=10^{2}$. Other parameter values are $k_1=2$, $k_2=3$, $\lambda=0.1$. The solid {magenta} curve in (b) represents the critical curve for which MLE $=0$, the regions below and above the critical curve denotes the unstable and stable synchronization state respectively. White region represents stable intralayer synchronization state.}\label{fig4}
\end{figure}
\par We have plotted $E_{intra} (\epsilon_1, \epsilon_2)$ in Fig. \ref{fig4} for slow and fast rewiring frequencies, respectively. A significant enhancement in synchronization region is noticed in Fig. \ref{fig4}(b) for sufficiently large rewiring frequency $(f=10^2)$ compared to the synchronization region for sufficiently slow rewiring frequency $(f=10^{-4})$, depicted in Fig. \ref{fig4}(a). For fast switching, we validate the numerical results through time-average network construction with theoretical predictions obtained from Eq.\eqref{eq.10}, similarly as previous example. The analytical threshold values for synchronization are portrayed in solid {magenta} line imposed to Fig. \ref{fig4}(b), which shows that the numerical simulation for fast switching case is in very good agreement with the analytical derivation for intralayer synchronization threshold.   

\begin{figure}[ht]
\centerline{\includegraphics[scale=0.4]{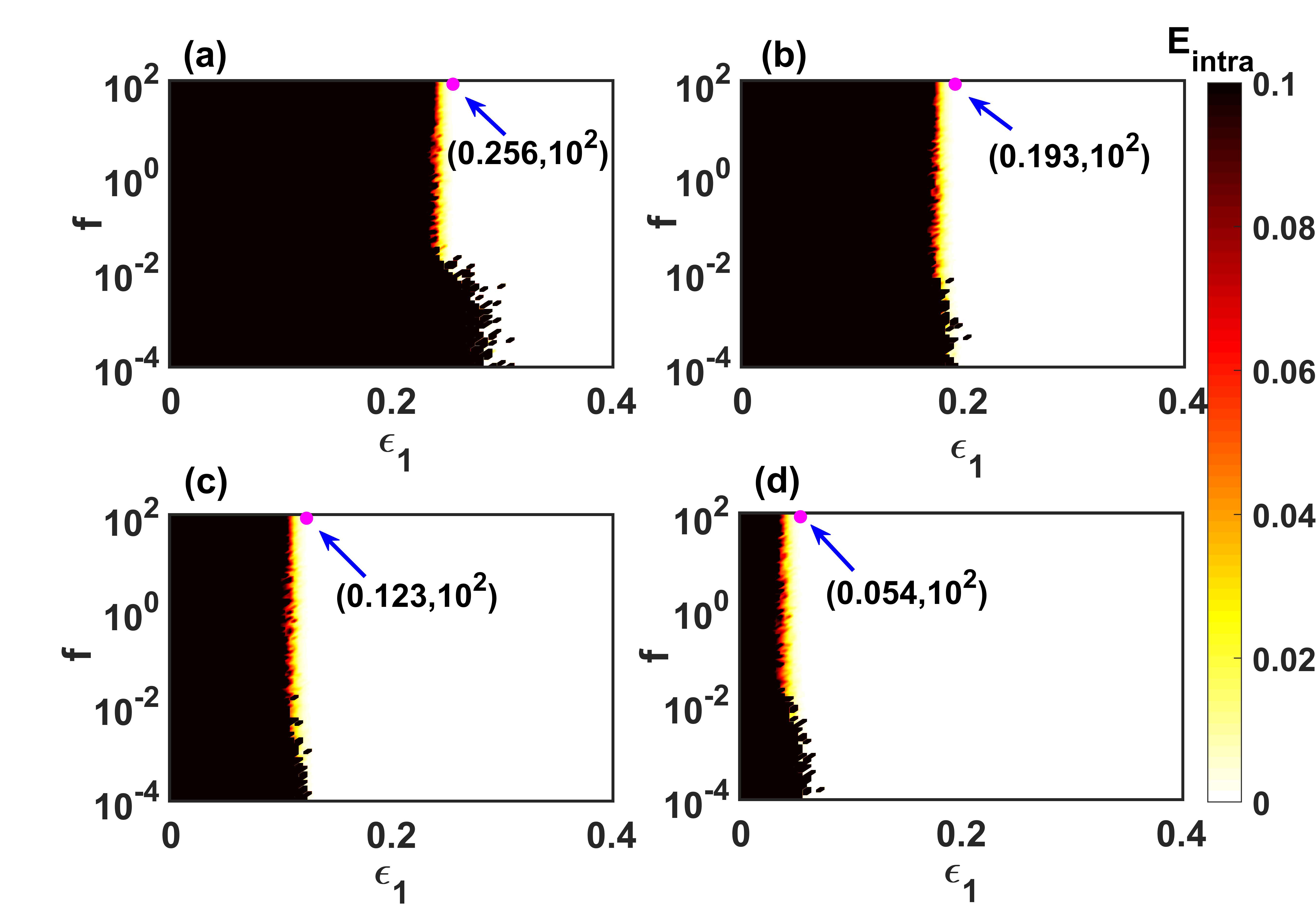}}
\caption{{\bf Coupled R\"{o}ssler oscillators:} Intralayer synchronization error in  ($\epsilon_{1}$,f) plane. For (a) $\epsilon_{2}=0.05$, (b)  $\epsilon_{2}=0.1$, (c)  $\epsilon_{2}=0.15$, and (d)  $\epsilon_{2}=0.2$. Other parameter values are $k_1=2=4$, $k_2=3$, and $\lambda=0.1$. {The solid magenta circles in each subfigure indicates analytically derived critical value of coupling strengths for sufficiently fast switching of intralayer network topologies. }}\label{fig5}
\end{figure}
\par Further, Fig. \ref{fig5} represents the region of coherence and incoherence in $(\epsilon_1,f)$ parameter plane for various values of tier-2 coupling strength $\epsilon_2$. For lower value of $\epsilon_2=0.05$, the synchronization region portrayed in Fig. \ref{fig5}(a). By increasing $\epsilon_2$ to $\epsilon_2=0.1$ the enhancement in synchrony region is shown in Fig. \ref{fig5}(b). Further increment of $\epsilon_2$ to $\epsilon_2=0.15$ (Fig. \ref{fig5}(c)) and $\epsilon_2=0.2$ (Fig. \ref{fig5}(d)) shows significant enhancement in synchronization region. In all this figure, the critical transition point for $\epsilon_1$ initially decreases as the rewiring frequency increases up to $f \approx 10^{-2}$. Beyond, $f=10^{-2}$ the critical point against $\epsilon_1$ is almost vertical. We found somewhat similar results as above when we plot the variation of synchronization error in $(\epsilon_2,f)$ plane for different values of $\epsilon_1$ (corresponding figures are not shown in the manuscript).

	\subsubsection{Coupled Sherman model of pancreatic beta cells interacting through  electrical and chemical synapses}
	
	We now scrutinize our framework to study the neuronal synchronization. Synchronization in neuronal networks is of enormous significance. Studies in neuroscience have identified existence of simultaneous interconnection between neurons through various synaptic transmissions, which can be perfectly schematized by multilayer networks \cite{sporns2010networks,bera2019intralayer}.  Here we consider ensemble of pancreatic beta cells, represented by paradigmatic Sherman model, interconnected concurrently through electrical and chemical synapses in a multilayer network framework. The dynamics of entire network is given by, 
	
	\begin{widetext}
	\begin{equation} \label{eq.18}
		\begin{array}{l}
			\mbox{Layer-1 :} \\
			\tau\dot{V}_{1i}=-I_{Ca}(V_{1i})-I_{K}(V_{1i},n_{1i})-I_{S}(V_{1i},s_{1i})+\epsilon_{1} \sum\limits_{j=1}^{N}{\mathscr{A}_{ij}^{[1,1]}}(t)(V_{1j}-V_{1i}) +\epsilon_{2}(E_{S}-V_{1i})\sum\limits_{j=1}^{N}{\mathscr{A}_{ij}^{[1,2]}}(t)\Gamma(V_{1j}) \\~~~~~~~~~~~~~ +\lambda(E_{S}-V_{1i})\Gamma(V_{2i}), \\
			\tau\dot{n}_{1i}=\mu [n^{\infty}(V_{1i})-n_{1i}], \\
			\tau_{s}\dot{s}_{1i}=s^{\infty}(V_{1i})-s_{1i}, \\\\
			
			\mbox{Layer-2 :} \\
			\tau\dot{V}_{2i}=-I_{Ca}(V_{2i})-I_{K}(V_{2i},n_{2i})-I_{S}(V_{2i},s_{2i})+\epsilon_{1} \sum\limits_{j=1}^{N}{\mathscr{A}_{ij}^{[2,1]}}(t)(V_{2j}-V_{2i})+\epsilon_{2}(E_{S}-V_{2i})\sum\limits_{j=1}^{N}{\mathscr{A}_{ij}^{[2,2]}}(t)\Gamma(V_{2j})
			\\~~~~~~~~~~~~~~~+\lambda(E_{S}-V_{2i})\Gamma(V_{1i}), \\
			\tau\dot{n}_{2i}=\mu[n^{\infty}(V_{2i})-n_{2i}], \\
			\tau_{s}\dot{s}_{2i}=s^{\infty}(V_{2i})-s_{2i},
		\end{array}
	\end{equation}
     \end{widetext}
	where $I_{Ca}(V)=g_{Ca}m^{\infty}(V-E_{Ca})$, $I_{K}(V,n)=g_{K}n(V-E_{K})$, $I_{S}(V,s)=g_{S}s(V-E_{K})$.
	$V_{li}$ is the membrane potential corresponding to the reversal potential $E_{Ca}=0.025$V, $E_{K}=-0.075$V. The time constants and maximum conductance are $\tau=0.02$, $\tau_{s}=5$, $g_{Ca}=3.6$, $g_{K}=10$, and $g_{S}=4$.  $\mu=1$, an auxiliary scaling factor, manages the time
	scale of the persistent potassium channels. The values of the gating variables at steady state are 
	\begin{equation*}
	\begin{array}{l}
		m^{\infty}(V)=\{1+\exp[-83.34(V+0.02)]\}^{-1}, \\n^{\infty}(V)=\{1+\exp[-178.57(V+0.016)]\}^{-1}, \;\mbox{and}\\ s^{\infty}(V)=\{1+\exp[-100(V+0.035345)]\}^{-1}.
	\end{array}	
	\end{equation*}
	
	\par We consider the neurons in a layer are interconnected simultaneously subject to electrical gap junctional coupling via electrical synapses and  chemical ion transportation via chemical synapses. The adjacency matrices corresponding to electrical synapses are $\mathscr{A}^{[l,1]}$, $l=1,2$, obtained from small-world network with average degree $k_{sw}$ and edge rewiring probability $p_{sw}$. $\mathscr{A}^{[l,2]}$, $l=1,2$ describes the structure of connections via chemical synapses, deliberated by random network with constant node degree $k_c$. Further, the neurons of two different layers are connected through chemical synapses.  The sigmoidal input-output function 
	\begin{eqnarray*}
		\Gamma(V)=\frac{1}{1+\exp(\lambda_{s}(V-\Theta_s))} ,
	\end{eqnarray*} describes the procedure for initiation of nonlinear chemical synapse. The synaptic reversal potential is $E_s=-0.02$. Slope of the sigmoidal function, and synaptic firing threshold are determined by the real-valued constants $\lambda_{s}$ and $\theta_{s}$, which are fixed at $\Theta_s=-0.045$ and $\lambda_{s}=-1000$. 

\begin{figure}[ht]
\centerline{\includegraphics[scale=0.45]{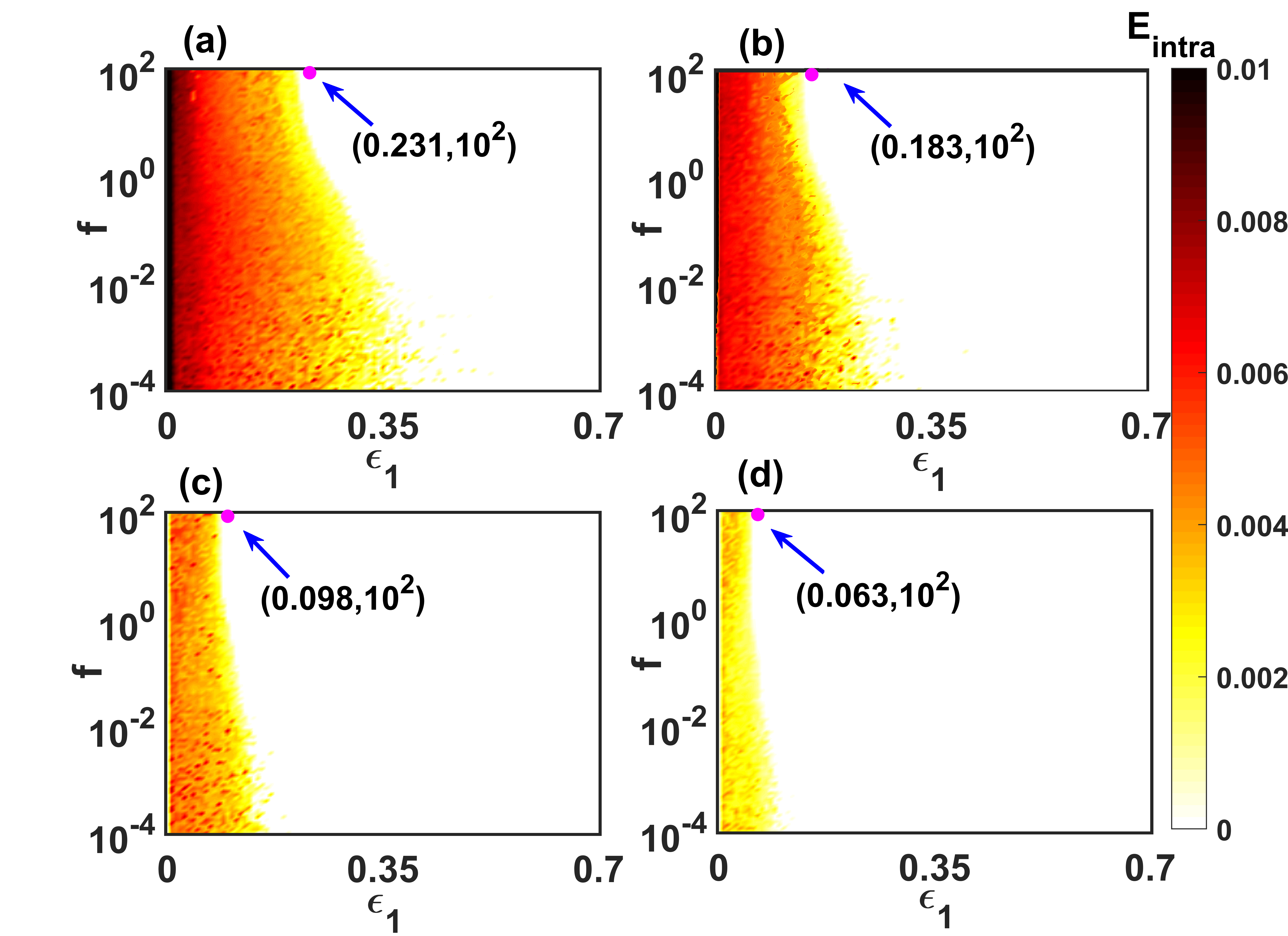}}
\caption{{\bf Coupled Sherman model:} Intralayer synchronization error in $(\epsilon_1,f)$ plane. For (a) $\lambda=0.0$, (b) $\lambda=0.02$, (c) $\lambda=0.04$, and (d) $\lambda=0.05$. Other parameter values are $\epsilon_2=0.015$, $p_{sw}=0.15$, $k_{sw}=4$, and $k_{c}=4$ respectively. {The solid magenta circles in each subfigure indicates analytically derived critical value of coupling strengths for sufficiently fast switching of intralayer network topologies. }}\label{fig6}
\end{figure}
\par
\begin{figure}[ht]
\centerline{\includegraphics[scale=0.44]{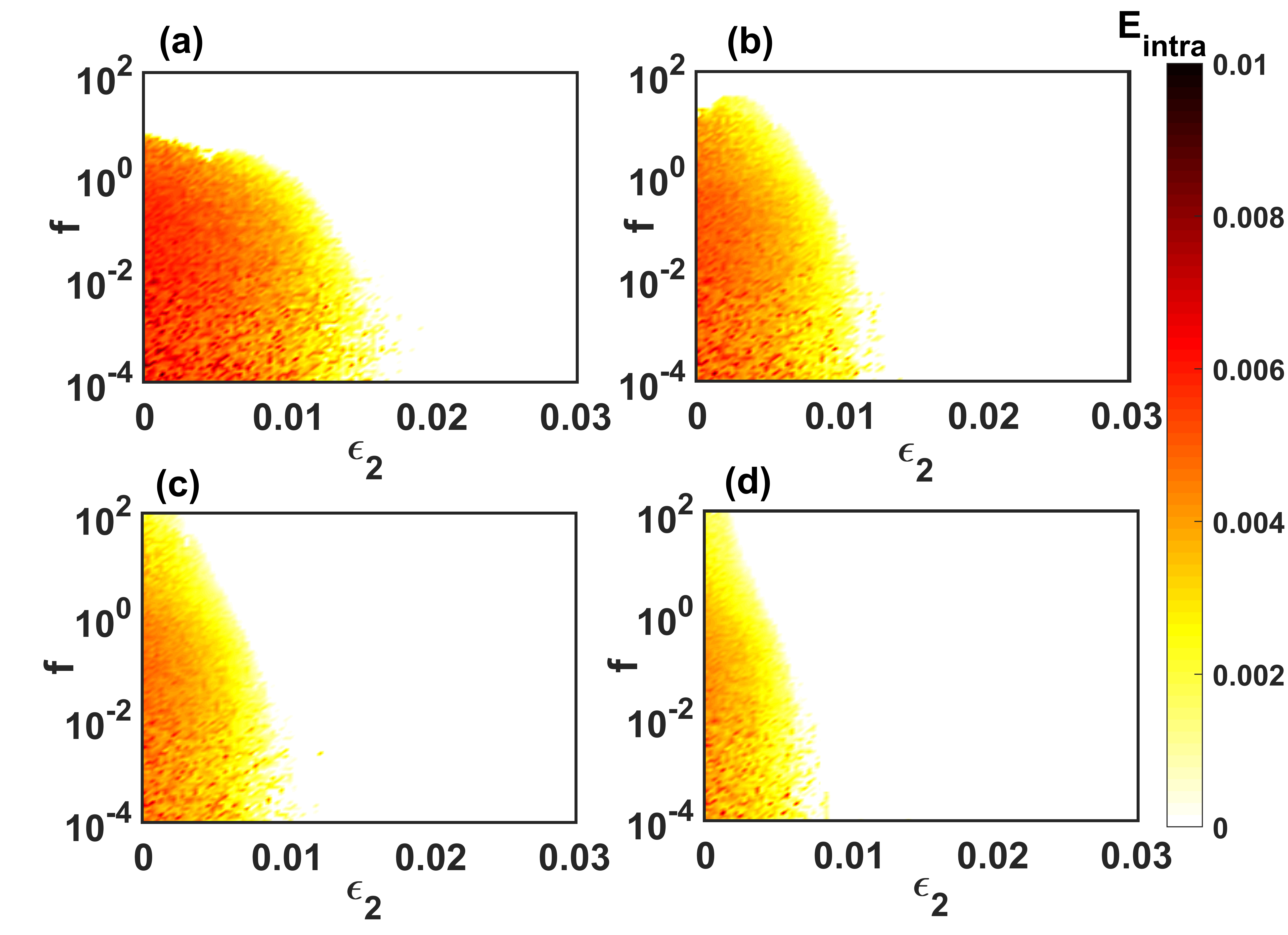}~~~~~~~~}
\caption{{\bf Coupled Sherman model:} Intralayer synchronization error in $(\epsilon_2,f)$ plane for (a) $\lambda=0.0$, (b) $\lambda=0.02$, (c) $\lambda=0.04$, and (d) $\lambda=0.05$. Other parameter values are $\epsilon_1=0.35$, $p_{sw}=0.15$, $k_{sw}=4$, and $k_{c}=4$ respectively.}\label{fig7}
\end{figure}

\begin{figure}[ht]
	\centerline{
		\includegraphics[scale=0.6]{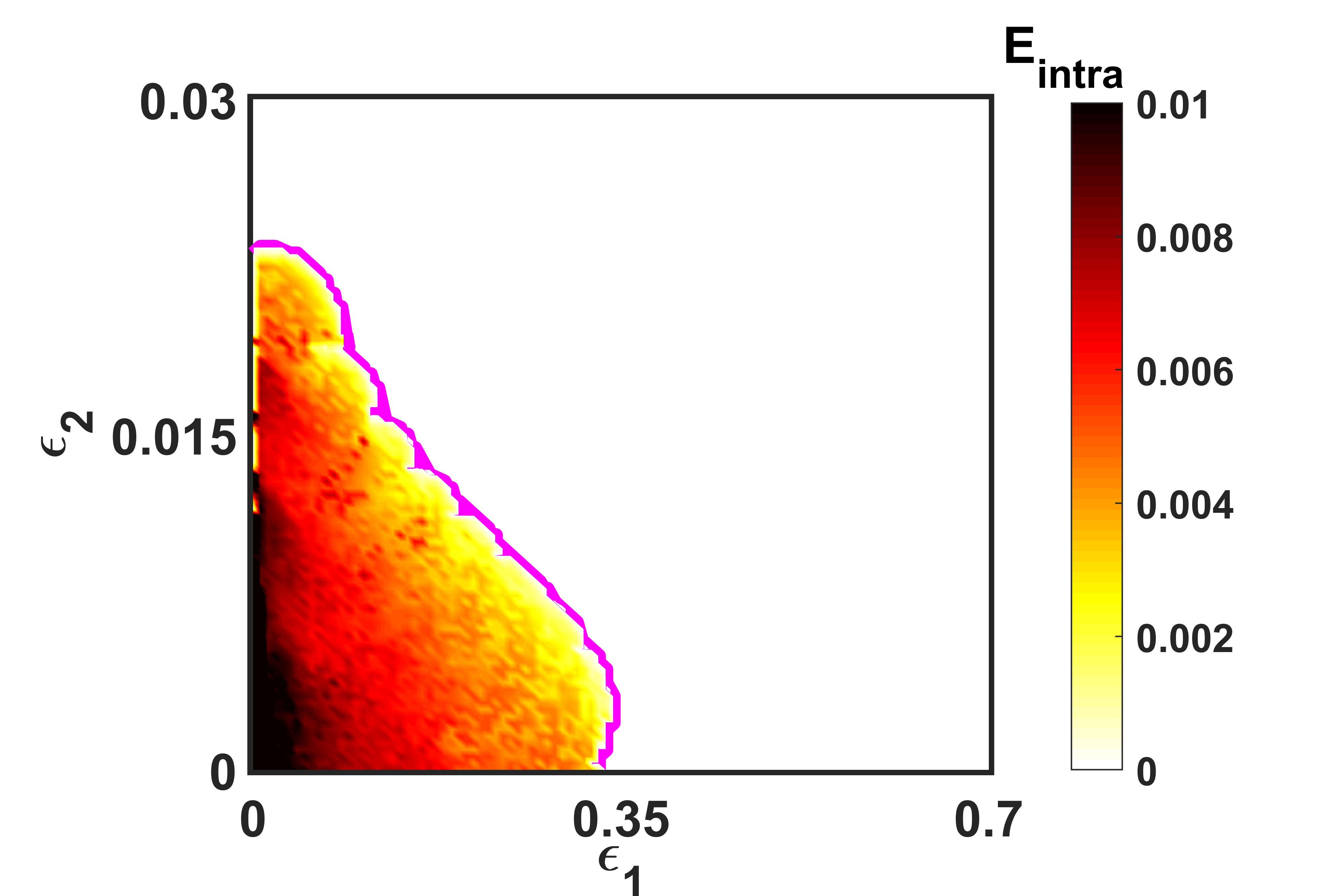}}
	
	\caption{  {\bf Coupled Sherman model:} Intralayer synchronization error in  $(\epsilon_1,\epsilon_2)$ parameter plane. The other parameter values are $\lambda=0.02$, $p_{sw}=0.15$, $k_{sw}=4$, and $k_c=4$. The solid {magenta} continuous line represents the curve corresponding to synchronization threshold MLE $=0$ which separates the synchrony and desynchrony region.}
	\label{fig8}
\end{figure}

 Figure \ref{fig6} portrays the result corresponding to intralayer synchronization in $(\epsilon_{1},f)$ parameter plane for various interlayer coupling strength $\lambda$. In the absence of interlayer coupling $(\lambda=0)$, the coherent and in-coherent regions are portrayed in Fig. \ref{fig6}(a) and by introducing $\lambda=0.02$, the enhancement in synchronization region is obtained in Fig. \ref{fig6}(b). Further increments of $\lambda$ to $\lambda=0.04$, and $0.05$, shows more enhancement, which are delineated in Figs. \ref{fig6}(c) and \ref{fig6}(d). Figures \ref{fig7}(a)-\ref{fig7}(d) represents similar results in $(\epsilon_{2},f)$ parameter plane for increasing value of interlayer coupling strength $\eta$. Apart from this, we notice that in neuronal network also the threshold to achieve synchronization lowers with the increase in rewiring frequency $f$. For sufficiently fast switching $(f=10^{2})$, $E_{intra} (\epsilon_1,\epsilon_2)$ is reported in Fig. \ref{fig8}, along with the analytical conjecture obtained from Eq.\eqref{eq.10} (shown in solid {magenta} line laid over the diagram), which confirms that the numerical result is in good agreement with our analytical derivation.

	\subsection{Numerical results of Simultaneous Block Diagonalization Method} \label{sbd_numerical}
	
\par In this subsection, we investigate the validity of our obtained theoretical results associated with simultaneous block diagonalization framework. For this, we consider the example of coupled Lorenz oscillators with sine coupling function described earlier in Eq.\eqref{eq.15}. The parameter values are taken same as before (See Sec. \ref{lorenz_numerical}). Each layer of the multilayer hypernetwork consists of $N=8$ nodes interconnected through two different tiers. The network architecture of each tier are taken as complete graph of N nodes by randomly removing two edges. In this scenario, we consider more systematically temporal networks that alternate between two different configurations. Particularly, the networks corresponding to each tier changes between two networks alternatively in odd and even time spans, i.e., each adjacency matrices $\mathscr{A}^{[1,1]}(t)$, $\mathscr{A}^{[1,2]}(t)$, $\mathscr{A}^{[2,1]}(t)$, and $\mathscr{A}^{[2,2]}(t)$ have two network realizations corresponding to odd and even times. The interlayer connections are all to all, i.e., one node in a layer is linked with each node of the other layer. 

\begin{figure}[ht]
\centerline{\includegraphics[scale=0.4]{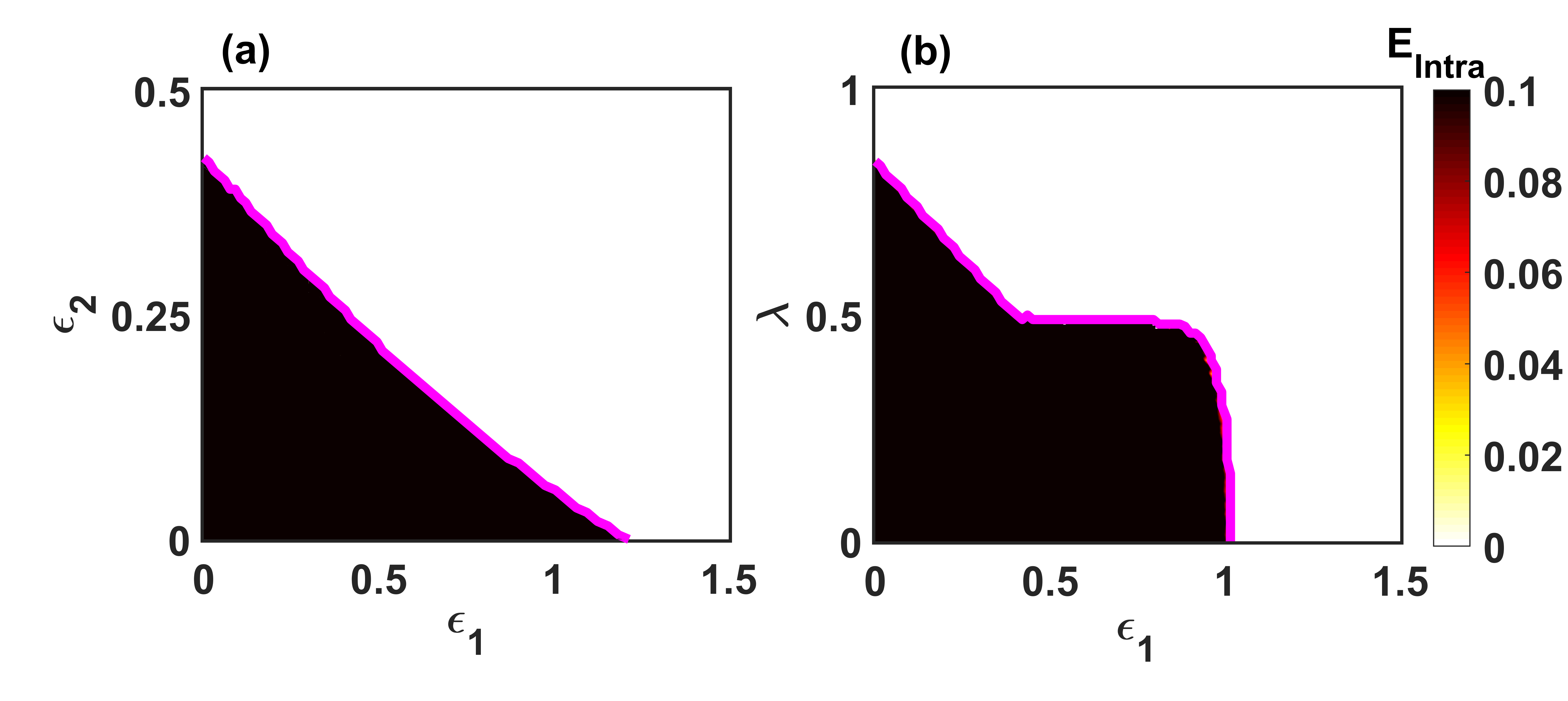}}
\caption{ Variation of intralayer synchronization error in (a) $(\epsilon_1,\epsilon_2)$, (b) $(\epsilon_1,\lambda)$ plane. The other parameter values are $\lambda=0.1$ for (a), and for (b) $\epsilon_2=0.05$, respectively. The solid {magenta} lines corresponds to the intralayer synchronization threshold obtained from analytical prediction.}\label{fig9}
\end{figure}
We first scrutinize the region of synchronization and desynchronization by evaluating the synchronization error $E_{intra}$. The corresponding results are portrayed in Fig. \ref{fig9}. Figure \ref{fig9}(a) describes $E_{intra}(\epsilon_1,\epsilon_2)$ for a fixed interlayer coupling strength $\lambda=0.1$. As $\epsilon_2$ increases the decreasing critical transition point of $\epsilon_1$ to achieve synchrony, confirms enhancement of intralayer synchronization. Figure \ref{fig9}(b) illustrates $E_{intra}(\epsilon_1, \lambda)$ for fixed intralayer coupling strength $\epsilon_2=0.05$ for tier-2. The synchronization threshold against $\epsilon_1$ is almost same with increasing $\lambda$ up to $\lambda \approx 0.5$. For $\lambda$ just greater than $0.5$, the critical point for $\epsilon_1$ suddenly decreases to $\epsilon_1=0.47$. Beyond that the critical coupling to achieve synchronization decreases as $\lambda$ increases. Similar phenomenon is also observed in $(\epsilon_{2},\lambda)$ parameter plane for fixed $\epsilon_1$ (corresponding figure is not shown here).

\begin{figure}[ht]
\centerline{\includegraphics[scale=0.85]{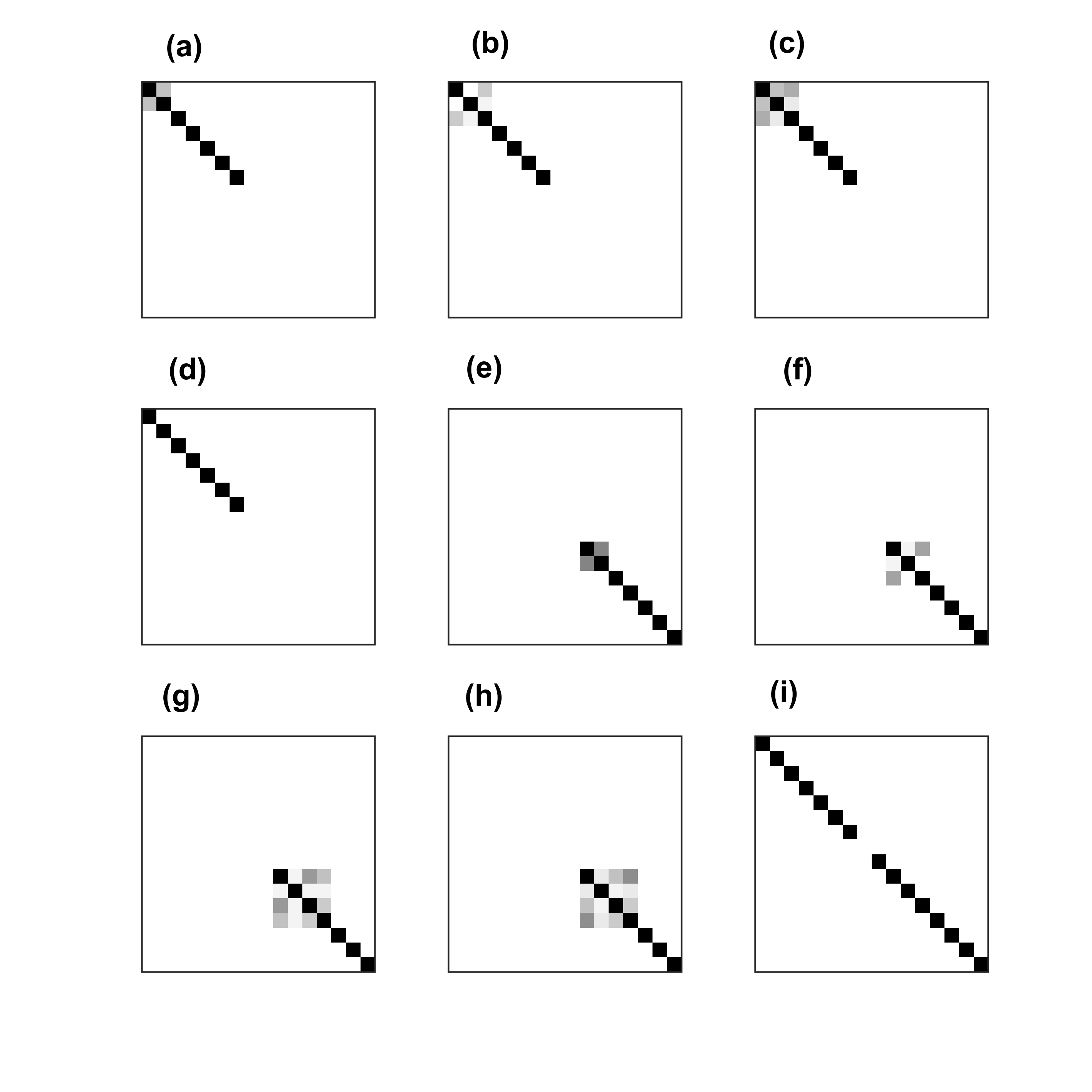}}
\caption{The block representations of all the non-commuting matrices in the set $S_{1}$ after applying SBD algorithm are portrayed from (a) to (i). Here the non-zero elements are colored in light to deep black according to their increasing values and the zero elements are colored in white.}\label{fig10}
\end{figure}
	
\par Next, we study the analytical stability of the synchronization state using SBD approach. As the network changes its configuration between even and odd times, the Laplacians associated with each tier also varies. Therefore, as discussed in Sec.\ref{sbd}, the variational equation \eqref{eq.12} will have a set of non-commuting matrices        $ \\S_1=\{U^{[1,1]}(\mbox{odd}), U^{[1,1]}(\mbox{even}),U^{[1,2]}(\mbox{odd}), U^{[1,2]}(\mbox{even}),\\\dots,U^{[2,2]}(\mbox{odd}), U^{[2,2]}(\mbox{even}),\mathscr{B}\}$. Here $U^{[l,\beta]}(\mbox{odd})$ and $U^{[l,\beta]}(\mbox{even})$, $l,\beta=1,2$ are associated with the network configuration at even and odd times. Applying SBD approach, we transform the above set of non-commuting matrices in common block diagonal form (Fig. \ref{fig10}). We notice that the set of matrices transforms into block diagonal matrices with blocks of sizes $1\times 1$, $2 \times 2$, $3 \times 3$ and $4 \times 4$. Among them two $1\times 1$ blocks are associated with the parallel modes to synchronization manifold and all the other blocks are associated with the transverse modes to synchronization manifold. Since, the maximum size of block is $4 \times 4$, we can rewrite the higher-dimensional transverse system in terms of lower-dimensional subsystems as, 
	\begin{equation} \label{eq.19}
		\begin{array}{l}
			\delta\dot{x}_{1i}= \sigma(\delta y_{1i}-\delta x_{1i}) -\epsilon_{1}\sum\limits_{j=1}^{4}{\mathscr{C}_{ij}^{[1,1]}(t)}\delta x_{1i} \\~~~~~~~~~~~~ -\lambda \mathscr{D}_{ii}^{[1]}\cos(x_2-x_1)\delta x_{1i},     \\
			
			\delta\dot{y}_{1i}=(\rho-z_1)\delta x_{1i}-\delta y_{1i} -x_{1}\delta z_{1i}-\epsilon_2 \sum\limits_{j=1}^{4}{\mathscr{C}_{ij}^{[1,2]}(t)} \delta y_{1i}, \\
			
			\delta\dot{z}_{1i}=y_{1}\delta x_{1i} +x_{1}\delta y_{1i}-\alpha \delta z_{1i}, \\\\
			
			\delta\dot{x}_{2i}= \sigma(\delta y_{2i}-\delta x_{2i}) -\epsilon_{1}\sum\limits_{j=1}^{4}{\mathscr{C}_{ij}^{[2,1]}(t)}\delta x_{2i} 
			\\~~~~~~~~~~~~ -\lambda \mathscr{D}_{ii}^{[2]} \cos(x_1-x_2)\delta x_{2i},     \\
			
			\delta\dot{y}_{2i}=(\rho-z_2)\delta x_{2i}-\delta y_{2i} -x_{2}\delta z_{2i}-\epsilon_2 \sum\limits_{j=1}^{4}{\mathscr{C}_{ij}^{[2,2]}(t)}\delta y_{2i}, \\
			
			\delta\dot{z}_{2i}=y_{2}\delta x_{2i} +x_{2}\delta y_{2i}-\alpha \delta z_{2i},    \hspace{10 pt} i=1,2,3,4.
			
		\end{array}  
	\end{equation}
	
	\begin{equation} \label{eq.20}
		\begin{array}{l}
			\delta\dot{x}_{1i}= \sigma(\delta y_{1i}-\delta x_{1i}) -\epsilon_{1}{\mathscr{C}_{ii}^{[1,1]}(t)}\delta x_{1i} \\ ~~~~~~~~~~ -\lambda \mathscr{D}_{ii}^{[1]}\cos(x_2-x_1)\delta x_{1i},     \\
			
			\delta\dot{y}_{1i}=(\rho-z_1)\delta x_{1i}-\delta y_{1i} -x_{1}\delta z_{1i}-\epsilon_2 {\mathscr{C}_{ii}^{[1,2]}(t)} \delta y_{1i}, \\
			
			\delta\dot{z}_{1i}=y_{1}\delta x_{1i} +x_{1}\delta y_{1i}-\alpha \delta z_{1i}, \\\\
			
			\delta\dot{x}_{2i}= \sigma(\delta y_{2i}-\delta x_{2i}) -\epsilon_{1}{\mathscr{C}_{ii}^{[2,1]}(t)}\delta x_{2i} \\ ~~~~~~~~~ -\lambda \mathscr{D}_{ii}^{[2]} \cos(x_1-x_2)\delta x_{2i}),     \\
			
			\delta\dot{y}_{2i}=(\rho-z_2)\delta x_{2i}-\delta y_{2i} -x_{2}\delta z_{2i}-\epsilon_2 {\mathscr{C}_{ii}^{[2,2]}(t)}\delta y_{2i}, \\
			
			\delta\dot{z}_{2i}=y_{2}\delta x_{2i} +x_{2}\delta y_{2i}-\alpha \delta z_{2i},   \hspace{10 pt} i=5,6,7
			
		\end{array}  
	\end{equation}
where $(x_1,y_1,z_1)$ and $(x_2,y_2,z_2)$ are the synchronized solutions. The matrices $\mathscr{C}^{[l,\beta]}(t)$, and $\mathscr{D}^{[l]}$, $l,\beta=1,2$ are the block diagonalized form of the intralayer Laplacians $\mathscr{L}^{[l,\beta]}(t)$ and the interlayer adjacencies $\mathscr{B}^{[1]},\mathscr{B}^{[2]}$, after discarding the $1\times1$ blocks associated with parallel mode. The maximum Lyapunov exponent corresponding to this lower-dimensional subsystems \eqref{eq.19}, \eqref{eq.20} gives the condition for stable intralayer coherent state of the temporal multiplex hypernetwork. In Figs. \ref{fig9}(a) - \ref{fig9}(c), the continuous {magenta} lines represent the curves corresponding to the synchronization threshold MLE $=0$. The regions below and above these curves respectively depict MLE $>0$ and MLE $<0$. Value of negative MLE gives stable synchronization state and positive MLE signifies unstable synchrony. These maximum Lyapunov exponent plots are found to be suitably confirmed with our numerical experiments.

\section{Conclusion}\label{conclusion}
\par Here we have considered a universal model attributing for temporal multilayer hypernetwork framework with arbitrary sorts of intralayer and interlayer coupling functions, and explicitly carried out the stability analysis of intralayer synchronous state. We derived the invariance condition of the coherent state and based on that condition look into the derivation of necessary condition for stable synchronization solution. The process of stability analysis is conducted by means of two well-known concepts - the fast switching approach and simultaneous block diagonalization technique. Employing fast switching approach, we show that the temporal network achieves stable intralayer synchronization state whenever corresponding static time-average structure is in stable intralayer coherent state. The condition for stable synchronous motion is derived with respect to time-averaged network structure, and we have delineated that, in some instances, our method resembles classical MSF approach. On the other hand, using simultaneous block diagonalization procedure, we derive the condition for stability of intralayer synchronous solution for temporal networks, without restricting to time-average network structure. In this case, we have shown that the transverse modes associated with synchronization manifold can be decoupled into finer form without any assumption. Finally, the analytical inferences are accompanied by several numerical results, which confirms the sustainability and universality of our approach. We expect, our study will pave the way for a new understanding of collective properties emerge in temporal multilayer framework. The use of our methods irrespective of coupling functions provide the scope to apply it for exploration of large class of coupling schemes associated with complex systems that can be represented by multilayer networks.

\appendix
\section{}
\subsection{Proof of resemblance between time-varying and time-averaged system}\label{time_average}

\par From Eq.\eqref{eq.1} and Eq.\eqref{eq.8}, one can easily obtain that the synchronization solution for both the time-varying and time-averaged network follows Eq.\eqref{eq.4}. Hence, without loss of generality, we assume that $\mathbf{\bar{x}}_i=\mathbf{x}_s$ and $\mathbf{\bar{y}}_i=\mathbf{y}_s$, for all $i=1,2,....,N$ at the state of synchrony. In order to analyze the stability of synchronization state for time-averaged case, we consider
$\delta \mathbf{\bar{x}}(t) =\bigl[ \delta \mathbf{\bar{x}}_1(t)^{tr}, \delta \mathbf{\bar{x}}_2(t)^{tr}, . . . ,\delta\mathbf{\bar{x}}_1(t)^{tr}\bigr]^{tr}$, and $\delta \mathbf{\bar{y}}(t) =\bigl[ \delta \mathbf{\bar{y}}_1(t)^{tr}, \delta \mathbf{\bar{y}}_2(t)^{tr}, . . . ,\delta\mathbf{\bar{y}}_1(t)^{tr}\bigr]^{tr}$, as small perturbations around the synchronization solution. Then the variational equation for time-averaged system can be expressed by,
\begin{widetext}
	\begin{equation}\label{eq.A1}
		\begin{array}{l}
			\delta\dot{{\bar{\mathbf{x}}}}=I_{N}{\otimes}JF_1(\mathbf{x}_s)\delta\mathbf{\bar{x}}+\sum\limits_{\beta=1}^{M}\epsilon_{\beta} d^{[\beta]}[I_N{\otimes}JG_{\beta}^{[1]}(\mathbf{x}_s,\mathbf{x}_s)+I_N{\otimes}DG_{\beta}^{[1]}(\mathbf{x}_s,\mathbf{x}_s)]\delta\mathbf{\bar{x}}  \\~~~~~~ -\sum\limits_{\beta=1}^{M}\epsilon_{\beta}\bar{\mathscr{L}}^{[\beta]}{\otimes}DG_{\beta}^{[1]}({\bf x}_s,{\bf x}_s)\delta\mathbf{\bar{x}}+\lambda[e^{[1]}I_{N}{\otimes}JH_1(\mathbf{x}_s,\mathbf{y}_s)\delta\mathbf{\bar{x}}+\mathscr{B}^{[1]}{\otimes}DH_1(\mathbf{x}_s,\mathbf{y}_s)\delta\mathbf{\bar{y}}], \\

			\delta\dot{{\bar{\mathbf{y}}}}=I_{N}{\otimes}JF_2(\mathbf{y}_s)\delta\mathbf{\bar{y}}+\sum\limits_{\beta=1}^{M}\epsilon_{\beta} d^{[\beta]}[I_N{\otimes}JG_{\beta}^{[2]}(\mathbf{y}_s,\mathbf{y}_s)+I_N{\otimes}DG_{\beta}^{[2]}(\mathbf{y}_s,\mathbf{y}_s)]\delta\mathbf{\bar{y}}  \\~~~~~~ -\sum\limits_{\beta=1}^{M}\epsilon_{\beta}\bar{\mathscr{L}}^{[\beta]}{\otimes}DG_{\beta}^{[1]}({\bf y}_s,{\bf y}_s)\delta\mathbf{\bar{y}}+\lambda[e^{[2]}I_{N}{\otimes}JH_2(\mathbf{y}_s,\mathbf{x}_s)\delta\mathbf{\bar{y}}+\mathscr{B}^{[2]}{\otimes}DH_2(\mathbf{y}_s,\mathbf{x}_s)\delta\mathbf{\bar{x}}].
			
		\end{array}
	\end{equation}
\end{widetext}

\par Since, each time-average Laplacians are real zero row-sum square matrices, all of their eigenvalues $\bar{\gamma}_{i}^{[\beta]} \in \mathbb{C}, i=1,2,...,N$ with one eigenvalue $\bar{\gamma}_{1}^{[\beta]}$ zero. Here, we have taken into consideration that the network structures corresponding to each tiers are connected. The associated set of eigenvectors forms an orthogonal basis $V^{[\beta]}$ of $\mathbb{C}^{N}$. As $\bar{\mathscr{L}}^{[\beta]}$ is real square matrix, it is unitarily triangularizable. So, there exists an upper triangular matrix $\bar{W}^{[\beta]}$ such that $\bar{W}^{[\beta]}={V^{[\beta]}}^{-1}{\bar{\mathscr{L}}}^{[\beta]}V^{[\beta]}$. We assume,
	\begin{equation*}
		\begin{array}{l}   	
			V^{[\beta]}= \begin{bmatrix}
				\frac{1}{\sqrt{N}} & v_{12}^{[\beta]} & v_{13}^{[\beta]} & ... & v_{1N}^{[\beta]} \\
				\frac{1}{\sqrt{N}} & v_{22}^{[\beta]} & v_{23}^{[\beta]} & ... & v_{2N}^{[\beta]} \\
				\\ 
				\frac{1}{\sqrt{N}} & v_{N2}^{[\beta]} & v_{N3}^{[\beta]} & ... & v_{NN}^{[\beta]} \\
			\end{bmatrix}
		\end{array},
	\end{equation*}
where the eigenvector corresponding to $\bar{\gamma}_{1}^{[\beta]} = 0$ is taken as the first column. The variational equation \eqref{eq.A1} contains all the parallel and transverse components to the synchronization manifold. To decouple the transverse modes from parallel one, we project the stack variables $\delta\mathbf{\bar{x}}$ and $\delta\mathbf{\bar{y}}$ onto the basis of eigenvectors $V^{[1]}$ corresponding to the time-averaged Laplacian of tier-1 by introducing new variables $\eta^{(\mathbf{\bar{x}})}=(V^{[1]}{\otimes}I_d)^{-1}\delta\mathbf{\bar{x}}$ and $\eta^{(\mathbf{\bar{y}})}=(V^{[1]}{\otimes}I_d)^{-1}\delta\mathbf{\bar{y}}$. The choice of the set of eigenvectors is completely arbitrary, one can choice any other basis and all the other eigenvector sets will eventually transform to such a basis through unitary matrix transformation. The generic equation \eqref{eq.A1} in terms of new variables then becomes, 

\begin{widetext}
\begin{equation}\label{eq.A2}
\begin{array}{l}
\dot{\eta}^{(\mathbf{\bar{x}})}=I_{N}{\otimes}JF_1(\mathbf{x}_s)\eta^{(\mathbf{\bar{x}})}+\sum\limits_{\beta=1}^{M}\epsilon_{\beta} d^{[\beta]}[I_N{\otimes}JG_{\beta}^{[1]}(\mathbf{x}_s,\mathbf{x}_s)+I_N{\otimes}DG_{\beta}^{[1]}(\mathbf{x}_s,\mathbf{x}_s)]\eta^{(\mathbf{\bar{x}})}  \\~~~~~~ -\sum\limits_{\beta=1}^{M}\epsilon_{\beta}({V^{[1]}}^{-1}\bar{\mathscr{L}}^{[\beta]}V^{[1]}){\otimes}DG_{\beta}^{[1]}({\bf x}_s,{\bf x}_s)\eta^{(\mathbf{\bar{x}})}+\lambda[e^{[1]}I_{N}{\otimes}JH_1(\mathbf{x}_s,\mathbf{y}_s)\eta^{(\mathbf{\bar{x}})}+({V^{[1]}}^{-1}\mathscr{B}^{[1]}V^{[1]}){\otimes}DH_1(\mathbf{x}_s,\mathbf{y}_s)\eta^{(\mathbf{\bar{y}})}], \\

\dot{\eta}^{(\mathbf{\bar{y}})}=I_{N}{\otimes}JF_2(\mathbf{y}_s)\eta^{(\mathbf{\bar{y}})}+\sum\limits_{\beta=1}^{M}\epsilon_{\beta} d^{[\beta]}[I_N{\otimes}JG_{\beta}^{[2]}(\mathbf{y}_s,\mathbf{y}_s)+I_N{\otimes}DG_{\beta}^{[2]}(\mathbf{y}_s,\mathbf{y}_s)]\eta^{(\mathbf{\bar{y}})}  \\~~~~~~ -\sum\limits_{\beta=1}^{M}\epsilon_{\beta}({V^{[1]}}^{-1}\bar{\mathscr{L}}^{[\beta]}V^{[1]}){\otimes}DG_{\beta}^{[1]}({\bf y}_s,{\bf y}_s)\eta^{(\mathbf{\bar{y}})}+\lambda[e^{[2]}I_{N}{\otimes}JH_2(\mathbf{y}_s,\mathbf{x}_s)\eta^{(\mathbf{\bar{y}})}+({V^{[1]}}^{-1}\mathscr{B}^{[2]}V^{[1]}){\otimes}DH_2(\mathbf{y}_o,\mathbf{x}_s)\eta^{(\mathbf{\bar{x}})}].
\end{array}
\end{equation}	
\end{widetext}

\par Using the triangularizable property of $\bar{\mathscr{L}}^{[\beta]}$ we have, 
	
	\begin{equation} {\label{eq.A3}}
		\begin{array}{l}
			{V^{[1]}}^{-1}\bar{\mathscr{L}}^{[\beta]}V^{[1]}= {V^{[1]}}^{-1}V^{[\beta]}\bar{U}^{[\beta]}{V^{[\beta]}}^{-1}V^{[1]}.
		\end{array}
	\end{equation}
	
	As the columns of ${V^{[\beta]}}$ are orthogonal eigenvectors, it is a unitary matrix, i.e., ${V^{[\beta]}}^{-1}={V^{[\beta]}}^{*}$, where ${^*}$ indicates conjugate transpose of a matrix. So, we can write,
	\begin{equation*}
		\begin{array}{l}
			{V^{[1]}}^{-1}V^{[\beta]}= \begin{bmatrix}
				1 & v_{12}^{[1,\beta]} & v_{13}^{[1,\beta]} & ... & v_{1N}^{[1,\beta]} \\
				0 & v_{22}^{[1,\beta]} & v_{23}^{[1,\beta]} & ... & v_{2N}^{[1,\beta]} \\
				\\ 
				0 & v_{N2}^{[1,\beta]} & v_{N3}^{[1,\beta]} & ... & v_{NN}^{[1,\beta]} \\
			\end{bmatrix},    
		\end{array}
	\end{equation*}
	
	and,\ 
	\begin{equation*} 
		\begin{array}{l}
			{V^{[\beta]}}^{-1}V^{[1]}= \begin{bmatrix}
				1 & v_{12}^{[\beta,1]} & v_{13}^{[\beta,1]} & ... & v_{1N}^{[\beta,1]} \\
				0 & v_{22}^{[\beta,1]} & v_{23}^{[\beta,1]} & ... & v_{2N}^{[\beta],1} \\
				\\ 
				0 & v_{N2}^{[\beta,1]} & v_{N3}^{[\beta,1]} & ... & v_{NN}^{[\beta,1]} \\
			\end{bmatrix}.    
		\end{array}
	\end{equation*}
	
	Substituting these expressions in \eqref{eq.A3}, we get
	\begin{equation}{\label{eq.A4}}
		\begin{array}{l}
			{V^{[1]}}^{-1}\bar{\mathscr{L}}^{[\beta]}V^{[1]}=\begin{bmatrix}
				0 & \bar{W}_1^{[\beta]} \\
				0_{(N-1) \times 1} & \bar{W}_2^{[\beta]}
			\end{bmatrix}	,
		\end{array}
	\end{equation}
	where $\bar{W}_1^{[\beta]}\in \mathbb{C}^{1\times (N-1)}$ and $ \bar{W}_2^{[\beta]} \in \mathbb{C}^{(N-1)\times (N-1)}$. 
	\par 
	Similarly, the triangularizable property of interlayer adjacency matrices and the identical node degrees for invariance of synchronization manifold gives,  
	\begin{equation}{\label{eq.A5}}
		\begin{array}{l}
			{V^{[1]}}^{-1}\mathscr{B}^{[l]}V^{[1]}=\begin{bmatrix}
				e^{[l]} & 0_{1\times (N-1)} \\
				0_{1\times (N-1)} & U_3^{[l]}
			\end{bmatrix}, \hspace{10 pt} l=1,2.
		\end{array}
	\end{equation}

\par Suppose, in parallel and transverse co-ordinates, the  transform variables $\eta^{\mathbf{(\bar{x})}}(t)$ and $\eta^{\mathbf{(\bar{y})}}(t)$ yield the decomposition $\eta^{\mathbf{(\bar{x})}}=[\eta_P^{\mathbf{(\bar{x})}},\eta_T^{\mathbf{(\bar{x})}}]$ and $\eta^{\mathbf{(\bar{y})}}=[\eta_P^{\mathbf{(\bar{y})}},\eta_T^{\mathbf{(\bar{y})}}]$ where $\eta_P^{\mathbf{(\bar{x})}},\eta_P^{\mathbf{(\bar{y})}} \in \mathbb{C}^d$ and $\eta_T^{\mathbf{(\bar{x})}},\eta_T^{\mathbf{(\bar{y})}} \in \mathbb{C}^{(N-1)d}$. Making these decomposition in \eqref{eq.A2} and substituting the expressions \eqref{eq.A4} and \eqref{eq.A5}, we get,  

\begin{widetext}
\begin{equation}\label{eq.A6}
\begin{array}{l}
\dot{\eta}_P^{(\mathbf{\bar{x}})}=JF_1(\mathbf{x}_s)\eta_P^{(\mathbf{\bar{x}})}+\sum\limits_{\beta=1}^{M}\epsilon_{\beta} d^{[\beta]}[JG_{\beta}^{[1]}(\mathbf{x}_s,\mathbf{x}_s)+DG_{\beta}^{[1]}(\mathbf{x}_s,\mathbf{x}_s)]\eta_P^{(\mathbf{\bar{x}})}  \\~~~~~~ -\sum\limits_{\beta=1}^{M}\epsilon_{\beta}\bar{W}_1^{[\beta]}{\otimes}DG_{\beta}^{[1]}({\bf x}_s,{\bf x}_s)\eta_T^{(\mathbf{\bar{x}})}+\lambda[e^{[1]}JH_1(\mathbf{x}_s,\mathbf{y}_s)\eta_P^{(\mathbf{\bar{x}})}+e^{[1]}DH_1(\mathbf{x}_s,\mathbf{y}_s)\eta_P^{(\mathbf{\bar{y}})}], \\

\dot{\eta}_T^{(\mathbf{\bar{x}})}=I_{N-1}{\otimes}JF_1(\mathbf{x}_s)\eta_T^{(\mathbf{\bar{x}})}+\sum\limits_{\beta=1}^{M}\epsilon_{\beta} d^{[\beta]}[I_{N-1}{\otimes}JG_{\beta}^{[1]}(\mathbf{x}_s,\mathbf{x}_s)+I_{N-1}{\otimes}DG_{\beta}^{[1]}(\mathbf{x}_s,\mathbf{x}_s)]\eta_T^{(\mathbf{\bar{x}})}  \\~~~~~~ -\sum\limits_{\beta=1}^{M}\epsilon_{\beta}\bar{W}_2^{[\beta]}{\otimes}DG_{\beta}^{[1]}({\bf x}_s,{\bf x}_s)\eta_T^{(\mathbf{\bar{x}})}+\lambda[e^{[1]}I_{N-1}{\otimes}JH_1(\mathbf{x}_s,\mathbf{y}_s)\eta_T^{(\mathbf{\bar{x}})}+U_3^{[1]}{\otimes}DH_1(\mathbf{x}_s,\mathbf{y}_s)\eta_T^{(\mathbf{\bar{y}})}],
\end{array}
\end{equation}
and,	
\begin{equation}\label{eq.A7}
\begin{array}{l}
\dot{\eta}_P^{(\mathbf{\bar{y}})}=JF_2(\mathbf{y}_s)\eta_P^{(\mathbf{\bar{y}})}+\sum\limits_{\beta=1}^{M}\epsilon_{\beta} d^{[\beta]}[JG_{\beta}^{[2]}(\mathbf{y}_s,\mathbf{y}_s)+DG_{\beta}^{[2]}(\mathbf{y}_s,\mathbf{y}_s)]\eta_P^{(\mathbf{\bar{y}})}  \\~~~~~~ -\sum\limits_{\beta=1}^{M}\epsilon_{\beta}\bar{W}_1^{[\beta]}{\otimes}DG_{\beta}^{[1]}({\bf y}_s,{\bf y}_s)\eta_T^{(\mathbf{\bar{y}})}+\lambda[e^{[2]}JH_2(\mathbf{y}_s,\mathbf{x}_s)\eta_P^{(\mathbf{\bar{y}})}+e^{[2]}DH_2(\mathbf{y}_s,\mathbf{x}_s)\eta_P^{(\mathbf{\bar{x}})}], \\

\dot{\eta}_T^{(\mathbf{\bar{y}})}=I_{N-1}{\otimes}JF_2(\mathbf{y}_s)\eta_T^{(\mathbf{\bar{y}})}+\sum\limits_{\beta=1}^{M}\epsilon_{\beta} d^{[\beta]}[I_{N-1}{\otimes}JG_{\beta}^{[2]}(\mathbf{y}_s,\mathbf{y}_s)+I_{N-1}{\otimes}DG_{\beta}^{[2]}(\mathbf{y}_s,\mathbf{y}_s)]\eta_T^{(\mathbf{\bar{y}})}  \\~~~~~~ -\sum\limits_{\beta=1}^{M}\epsilon_{\beta}\bar{W}_2^{[\beta]}{\otimes}DG_{\beta}^{[2]}({\bf y}_s,{\bf y}_s)\eta_T^{(\mathbf{\bar{y}})}+\lambda[e^{[2]}I_{N-1}{\otimes}JH_2(\mathbf{y}_s,\mathbf{x}_s)\eta_T^{(\mathbf{\bar{y}})}+U_3^{[2]}{\otimes}DH_2(\mathbf{y}_s,\mathbf{x}_s)\eta_T^{(\mathbf{\bar{x}})}].
\end{array}
\end{equation}
\end{widetext}

Notably, the evolution of latter components, i.e., $\eta_T^{(\mathbf{\bar{x}})}$ and $\eta_T^{(\mathbf{\bar{y}})}$ are independent of the former ones, i.e., $\eta_P^{(\mathbf{\bar{x}})}$ and $\eta_P^{(\mathbf{\bar{y}})}$. The dynamics of the former ones, attributing to the motion parallel along the intralayer synchronization manifold, and that for the latter ones characterizing the transverse modes to the synchronization state.  
	\par Assuming $\zeta_d(t)=[{\eta_T^{\mathbf{(\bar{x})}}}^{tr}  {\eta_T^{\mathbf{(\bar{y})}}}^{tr} ]^{tr} \in \mathbb{C}^{2(N-1)d}$, the evolution of transverse modes are rewritten as, 
\begin{widetext}
	\begin{equation}{\label{eq.A8}}
		\begin{array}{l}
			\dot{\zeta}_d(t)=\big[ A(t)-\sum\limits_{\beta=1}^{M}{\epsilon_{\beta}}(\bar{E}_1^{[\beta]}\otimes DG_{\beta}^{[1]}(\mathbf{x}_s,\mathbf{x}_s)+\bar{E}_2^{[\beta]}\otimes DG_{\beta}^{[2]}(\mathbf{y}_s,\mathbf{y}_s)) \big] \zeta_d(t),
		\end{array}
	\end{equation}
where
\begin{equation*} 
\begin{array}{l}
A(t)=\begin{bmatrix}
A_{11}(t) & \lambda U_3^{[1]}{\otimes}DH_1(\mathbf{x}_s,\mathbf{y}_s) \\
\lambda U_3^{[2]}{\otimes}DH_2(\mathbf{y}_s,\mathbf{x}_s) & A_{22}(t)
\end{bmatrix},
\end{array}
\end{equation*}
\begin{equation*} 
\begin{array}{lll}
A_{11}(t)=I_{N-1}{\otimes}JF_1(\mathbf{x}_s)+\sum\limits_{\beta=1}^{M}\epsilon_{\beta} d^{[\beta]}[I_{N-1}{\otimes}JG_{\beta}^{[1]}(\mathbf{x}_s,\mathbf{x}_s)+I_{N-1}{\otimes}DG_{\beta}^{[1]}(\mathbf{x}_s,\mathbf{x}_s)]+\lambda e^{[1]}I_{N-1}{\otimes}JH_1(\mathbf{x}_s,\mathbf{y}_s), \\

A_{22}(t)=I_{N-1}{\otimes}JF_2(\mathbf{y}_s)+\sum\limits_{\beta=1}^{M}\epsilon_{\beta} d^{[\beta]}[I_{N-1}{\otimes}JG_{\beta}^{[2]}(\mathbf{y}_s,\mathbf{y}_s)+I_{N-1}{\otimes}DG_{\beta}^{[2]}(\mathbf{y}_s,\mathbf{y}_s)]+\lambda e^{[2]}I_{N-1}{\otimes}JH_2(\mathbf{y}_s,\mathbf{x}_s),
\end{array}
\end{equation*}
\end{widetext}
and,
\begin{equation*}
		\begin{array}{l}
			\bar{E}_1^{[\beta]}=\begin{bmatrix}
				\bar{W}_2^{[\beta]} & 0_{(N-1) \times (N-1)} \\
				0_{(N-1) \times (N-1)} & 0_{(N-1) \times (N-1)}
			\end{bmatrix},~\mbox{and}
				\end{array}
	\end{equation*}

\begin{equation*}
	\begin{array}{l}
			\bar{E}_2^{[\beta]}=\begin{bmatrix}
				0_{(N-1) \times (N-1)} & 0_{(N-1) \times (N-1)} \\
				0_{(N-1) \times (N-1)} & \bar{W}_2^{[\beta]}
			\end{bmatrix}.
		\end{array}
	\end{equation*}
	
	\par
	Now, we implement the identical co-ordinate transformation to the variational equation \eqref{eq.5} corresponding to the temporal multilayer hypernetwork by projecting the stack variables on the same basis of eigenvectors $V^{[1]}$. The new variables are then defined as, 
	\begin{equation*}
		\begin{array}{l}
			\eta^{(\mathbf{x})}=(V^{[1]}{\otimes}I_d)^{-1}\delta\mathbf{x} \;\; \mbox{and} \;\; \eta^{(\mathbf{y})}=(V^{[1]}{\otimes}I_d)^{-1}\delta\mathbf{y}.
		\end{array}
	\end{equation*} Proceeding as earlier and by spectral decomposition of stack variables into parallel and transverse modes, we can express the evolution of variational equation transverse to synchronization manifold as,
\begin{widetext}
	\begin{equation}{\label{eq.A9}}
		\begin{array}{l}
			\dot{\zeta}_e(t)=[A(t)-\sum\limits_{\beta=1}^{M}{\epsilon_{\beta}}({E}_1^{[\beta]}(t)\otimes DG_{\beta}^{[1]}(\mathbf{x}_s,\mathbf{x}_s)+{E}_2^{[\beta]}(t)\otimes DG_{\beta}^{[2]}(\mathbf{y}_s,\mathbf{y}_s))] \zeta_e(t),
		\end{array}
	\end{equation}
\end{widetext}
	where $\zeta_e(t)=[{\eta_T^{\mathbf{(x)}}}^{tr}  {\eta_T^{\mathbf{(y)}}}^{tr} ]^{tr} \in \mathbb{C}^{2(N-1)d}$ represents the state of the transverse mode, and 
	\begin{equation*}
		\begin{array}{l}
			{E}_1^{[\beta]}(t)=\begin{bmatrix}
				{W}_2^{[1,\beta]}(t) & 0_{(N-1) \times (N-1)} \\
				0_{(N-1) \times (N-1)} & 0_{(N-1) \times (N-1)}
			\end{bmatrix},
			\;\;\;\mbox{and} \\\\
			{E}_2^{[\beta]}(t)=\begin{bmatrix}
				0_{(N-1) \times (N-1)} & 0_{(N-1) \times (N-1)} \\
				0_{(N-1) \times (N-1)} & {W}_2^{[2,\beta]}(t)
			\end{bmatrix}.
		\end{array}
	\end{equation*}
	${W}_2^{[l,\beta]}(t)$, $l=1,2$ comes from the fact that time-varying Laplacians are similar to a upper triangular matrix ${W}^{[l,\beta]}(t) = \begin{bmatrix}
		0 & {W}_1^{[l,\beta]}\\
		0_{(N-1)\times 1} & {W}_2^{[l,\beta]}
	\end{bmatrix}$, such that ${{V^{[1]}}^{-1}{\mathscr{L}}^{[l,\beta]}(t)V^{[1]}}= {W}^{[l,\beta]}(t)$.	
	
	\par 
	Assuming, $\bar{\mathscr{L}}^{[\beta]}=\frac{1}{T} \int_{t}^{t+T}{\mathscr{L}^{[l,\beta]}(z)} dz$, $\beta=1,2,...,M ; l=1,2$ gives, 
	
	\begin{equation*}
		\begin{array}{l}
			{V^{[1]}}^{-1}\bar{\mathscr{L}}^{[\beta]}V^{[1]}=\frac{1}{T}\int_{t}^{t+T}{{V^{[1]}}^{-1}{\mathscr{L}}^{[l,\beta]}(z)V^{[1]}} dz.
		\end{array}
	\end{equation*}
	
	From this interpretation one can easily deduce that, 
	\begin{equation*}
		\begin{array}{l}
			\bar{W}_2^{[\beta]}=\frac{1}{T} \int_{t}^{t+T}{W_2^{[l,\beta]}(z)} dz,
		\end{array}
	\end{equation*} which implies $\bar{E}_l^{[\beta]}=\frac{1}{T} \int_{t}^{t+T}{E_l^{[\beta]}(z)} dz$ for $l=1,2$.
	
	Thus, with the reference of Lemma \ref{lemma_1}, we come to a conclusion that the transverse error system corresponding to time-varying structure (Eq.\eqref{eq.A9}) stabilizes whenever the all transverse modes of the time-averaged system (given by Eq.\eqref{eq.A8}) extinct. The extinction of all transverse modes indicates a stable synchronization state. Hence, the stable intralayer coherent state of the time-averaged multilayer hypernetwork \eqref{eq.8} implies achievement of stable intralayer synchrony for time-varying network \eqref{eq.1}.

	\subsection{Derivation of Eq.\eqref{eq.10}}\label{commutative}
	Since all the time average Laplacians and interlayer adjacency matrices are real and symmetric, they are all diagonalizable.
	Let $\bar{\mathscr{L}}^{[1]}$ commutes with all the other time average Laplacian matrices $\bar{\mathscr{L}}^{[\beta]},\;(\beta=2,3,...,M)$ as well as interlayer adjacency matrices $\mathscr{B}^{[l]},\;l=1,2$. Since, all the above matrices are diagonalizable and $\bar{\mathscr{L}}^{[1]}$ commutes with all other matrices, they share a common basis of eigenvectors and $V^{[1]}$ diagonalizes $\bar{\mathscr{L}}^{[1]}$. Then $V^{[1]}$ also diagonalizes all the other matrices. Therefore, \begin{equation*}
		\begin{array}{l}
			{V^{[1]}}^{-1}\bar{\mathscr{L}}^{[\beta]}V^{[1]} = diag\{0<\bar{\gamma}_{2}^{[\beta]}\le\bar{\gamma}_{3}^{[\beta]},...,\le\bar{\gamma}_{N}^{[\beta]}\},
		\end{array}
	\end{equation*}
	and 
\begin{equation*}
\begin{array}{l}
{V^{[1]}}^{-1}{\mathscr{B}}^{[l]}V^{[1]} = diag\{e^{[l]}=\varGamma_{1}^{[l]},\varGamma_{2}^{[l]},\varGamma_{3}^{[l]},\dots,\varGamma_{N}^{[l]}\}.
\end{array}
\end{equation*}
Here $\bar{\gamma}_{i}^{[\beta]}$ and $\varGamma_{i}^{[l]}$ for $i=1,2,\dots,N$ are the eigenvalues of time-averaged Laplacian matrices corresponding to tier-$\beta$ and that of interlayer adjacency matrices. This implies $\bar{W}_2^{[\beta]}=diag\{\gamma_{2}^{[\beta]}\le\gamma_{3}^{[\beta]},...,\le\gamma_{N}^{[\beta]}\}$, and $U_{3}^{[l]}=diag\{\varGamma_{2}^{[l]},\varGamma_{3}^{[l]},\dots,\varGamma_{N}^{[l]}\}.$ Hence, by substituting the above expressions in the transverse error system \eqref{eq.9} or, \eqref{eq.A8}, we have
\begin{widetext}
\begin{equation}\label{eq.A10}
\begin{array}{l}
\dot{\eta}_{T_i}^{(\mathbf{\bar{x}})}=JF_1(\mathbf{x}_s)\eta_{T_i}^{(\mathbf{\bar{x}})}+\sum\limits_{\beta=1}^{M}\epsilon_{\beta} d^{[\beta]}[JG_{\beta}^{[1]}(\mathbf{x}_s,\mathbf{x}_s)+DG_{\beta}^{[1]}(\mathbf{x}_s,\mathbf{x}_s)]\eta_{T_i}^{(\mathbf{\bar{x}})}  \\~~~~~~ -\sum\limits_{\beta=1}^{M}\epsilon_{\beta}\bar{\gamma}_{i}^{[\beta]}DG_{\beta}^{[1]}({\bf x}_s,{\bf x}_s)\eta_{T_i}^{(\mathbf{\bar{x}})}+\lambda[e^{[1]}JH_1(\mathbf{x}_s,\mathbf{y}_s)\eta_{T_i}^{(\mathbf{\bar{x}})}+\varGamma_{i}^{[1]}DH_1(\mathbf{x}_s,\mathbf{y}_s)\eta_{T_i}^{(\mathbf{\bar{y}})}], \\
\dot{\eta}_{T_i}^{(\mathbf{\bar{y}})}=JF_2(\mathbf{y}_s)\eta_{T_i}^{(\mathbf{\bar{y}})}+\sum\limits_{\beta=1}^{M}\epsilon_{\beta} d^{[\beta]}[JG_{\beta}^{[2]}(\mathbf{y}_s,\mathbf{y}_s)+DG_{\beta}^{[2]}(\mathbf{y}_s,\mathbf{y}_s)]\eta_{T_i}^{(\mathbf{\bar{y}})}  \\~~~~~~ -\sum\limits_{\beta=1}^{M}\epsilon_{\beta}\bar{\gamma}_{i}^{[\beta]}DG_{\beta}^{[2]}({\bf y}_s,{\bf y}_s)\eta_{T_i}^{(\mathbf{\bar{y}})}+\lambda[e^{[2]}JH_2(\mathbf{y}_s,\mathbf{x}_s)\eta_{T_i}^{(\mathbf{\bar{y}})}+\varGamma_{i}^{[2]}DH_2(\mathbf{y}_s,\mathbf{x}_s)\eta_{T_i}^{(\mathbf{\bar{x}})}], 
\end{array}
\end{equation}
\end{widetext}
$i=2,3.\dots,N$. Eq.\eqref{eq.A10} clearly shows that the $2d(N-1)$-dimensional coupled transverse error symstem is now decoupled into $(N-1)$ subsystems, each are of $2d$ dimension.

\subsection{Dimensionality reduction for single tier}\label{one_tier}
\par As we have assumed that each layer of the multilayer hypernetwork has only one tier, let the corresponding time average Laplacian matrix $\bar{\mathscr{L}}^{[1]}$ with the set of eigenvalues $\{0,\underbrace{\gamma,\dots,\gamma}_{(N-1) times}\}$. Therefore, ${V^{[1]}}^{-1}\bar{\mathscr{L}}^{[1]}V^{[1]}=\{0,\underbrace{\gamma,\dots,\gamma}_{(N-1) times}\}$. It implies, $\bar{W}_2^{[1]} = diag\{\underbrace{\gamma,\dots,\gamma}_{(N-1) times}\}$. Also, by our assumption, $\mathscr{B}^{[1]}=\mathscr{B}^{[2]}=\mathscr{B}$. So, we have, $e^{[1]}=e^{[2]}=e (say)$ and $U_3^{[1]}=U_3^{[2]}=U_3 (say)$. Then the transversed error equation \eqref{eq.9} or, \eqref{eq.A8} becomes,
\begin{widetext}
\begin{equation}\label{eq.A11}
\begin{array}{l}
\dot{\eta}_T^{(\mathbf{\bar{x}})}=I_{N-1}{\otimes}JF_1(\mathbf{x}_s)\eta_T^{(\mathbf{\bar{x}})}+\epsilon_{1} d^{[1]]}[I_{N-1}{\otimes}JG_{1}^{[1]}(\mathbf{x}_s,\mathbf{x}_s)+I_{N-1}{\otimes}DG_{1}^{[1]}(\mathbf{x}_s,\mathbf{x}_s)]\eta_T^{(\mathbf{\bar{x}})}  \\~~~~~~ -\epsilon_{1}\gamma DG_{1}^{[1]}({\bf x}_s,{\bf x}_s)\eta_T^{(\mathbf{\bar{x}})}+\lambda[e I_{N-1}{\otimes}JH_1(\mathbf{x}_s,\mathbf{y}_s)\eta_T^{(\mathbf{\bar{x}})}+U_3{\otimes}DH_1(\mathbf{x}_s,\mathbf{y}_s)\eta_T^{(\mathbf{\bar{y}})}],\\\\

\dot{\eta}_T^{(\mathbf{\bar{y}})}=I_{N-1}{\otimes}JF_2(\mathbf{y}_s)\eta_T^{(\mathbf{\bar{y}})}+\epsilon_{1} d^{1}[I_{N-1}{\otimes}JG_{1}^{[2]}(\mathbf{y}_s,\mathbf{y}_s)+I_{N-1}{\otimes}DG_{1}^{[2]}(\mathbf{y}_s,\mathbf{y}_s)]\eta_T^{(\mathbf{\bar{y}})}  \\~~~~~~ -\epsilon_{1}\gamma DG_{1}^{[2]}({\bf y}_s,{\bf y}_s)\eta_T^{(\mathbf{\bar{y}})}+\lambda[e I_{N-1}{\otimes}JH_2(\mathbf{y}_s,\mathbf{x}_s)\eta_T^{(\mathbf{\bar{y}})}+U_3{\otimes}DH_2(\mathbf{y}_s,\mathbf{x}_s)\eta_T^{(\mathbf{\bar{x}})}].
\end{array}
\end{equation}
\end{widetext}
Every terms of the above transverse error equations are block diagonalized except $U_3\otimes DH_1({\bf x}_s,{\bf y}_s)$ and $U_3\otimes DH_2({\bf y}_s,{\bf x}_s)$.  Now since the interlayer adjacency matrix is symmetric, $U_3$ is also symmetric. Hence it is diagonalizable by its basis of eigenvector. Now if the eigenvalues of the interlayer adjacency matrix $\mathscr{B}$ are $\{e,\varGamma_{2},\varGamma_{3},...,\varGamma_{N}\}$, then projecting the error system on the eigenspace of $U_3$ say, $V_3$ we get ,
	\begin{equation*}
		\begin{array}{l}
			V_3^{-1}U_3V_3= \mbox{diag}\{\varGamma_{2},\varGamma_{3},...,\varGamma_{N}\}.
		\end{array}
	\end{equation*}
	Hence Eq.\eqref{eq.A11} can be rewritten as,

\begin{widetext}
\begin{equation}\label{eq.A12}
\begin{array}{lll}
\dot{\eta}_{T_i}^{(\mathbf{\bar{x}})}=JF_1(\mathbf{x}_s)\eta_{T_i}^{(\mathbf{\bar{x}})}+\epsilon_{1} d^{[1]]}[JG_{1}^{[1]}(\mathbf{x}_s,\mathbf{x}_s)+DG_{1}^{[1]}(\mathbf{x}_s,\mathbf{x}_s)]\eta_{T_i}^{(\mathbf{\bar{x}})}  \\~~~~~~ -\epsilon_{1}\gamma DG_{1}^{[1]}({\bf x}_s,{\bf x}_s)\eta_{T_i}^{(\mathbf{\bar{x}})}+\lambda[e JH_1(\mathbf{x}_s,\mathbf{y}_s)\eta_{T_i}^{(\mathbf{\bar{x}})}+\varGamma_{i}DH_1(\mathbf{x}_s,\mathbf{y}_s)\eta_{T_i}^{(\mathbf{\bar{y}})}],\\\\

\dot{\eta}_{T_i}^{(\mathbf{\bar{y}})}=JF_2(\mathbf{y}_s)\eta_{T_i}^{(\mathbf{\bar{y}})}+\epsilon_{1} d^{1}[JG_{1}^{[2]}(\mathbf{y}_s,\mathbf{y}_s)+DG_{1}^{[2]}(\mathbf{y}_s,\mathbf{y}_s)]\eta_{T_i}^{(\mathbf{\bar{y}})}  \\~~~~~~ -\epsilon_{1}\gamma DG_{1}^{[2]}({\bf y}_s,{\bf y}_s)\eta_{T_i}^{(\mathbf{\bar{y}})}+\lambda[e JH_2(\mathbf{y}_s,\mathbf{x}_s)\eta_{T_i}^{(\mathbf{\bar{y}})}+\varGamma_{i}DH_2(\mathbf{y}_s,\mathbf{x}_s)\eta_{T_i}^{(\mathbf{\bar{x}})}],
\end{array}
\end{equation}
\end{widetext} 
$i=2,3,\dots,N$. This is certainly a N-1 number of $2d$-dimensional equations.

\bibliographystyle{apsrev4-1} 

\begin{thebibliography}{51}%
	\makeatletter
	\providecommand \@ifxundefined [1]{%
		\@ifx{#1\undefined}
	}%
	\providecommand \@ifnum [1]{%
		\ifnum #1\expandafter \@firstoftwo
		\else \expandafter \@secondoftwo
		\fi
	}%
	\providecommand \@ifx [1]{%
		\ifx #1\expandafter \@firstoftwo
		\else \expandafter \@secondoftwo
		\fi
	}%
	\providecommand \natexlab [1]{#1}%
	\providecommand \enquote  [1]{``#1''}%
	\providecommand \bibnamefont  [1]{#1}%
	\providecommand \bibfnamefont [1]{#1}%
	\providecommand \citenamefont [1]{#1}%
	\providecommand \href@noop [0]{\@secondoftwo}%
	\providecommand \href [0]{\begingroup \@sanitize@url \@href}%
	\providecommand \@href[1]{\@@startlink{#1}\@@href}%
	\providecommand \@@href[1]{\endgroup#1\@@endlink}%
	\providecommand \@sanitize@url [0]{\catcode `\\12\catcode `\$12\catcode
		`\&12\catcode `\#12\catcode `\^12\catcode `\_12\catcode `\%12\relax}%
	\providecommand \@@startlink[1]{}%
	\providecommand \@@endlink[0]{}%
	\providecommand \url  [0]{\begingroup\@sanitize@url \@url }%
	\providecommand \@url [1]{\endgroup\@href {#1}{\urlprefix }}%
	\providecommand \urlprefix  [0]{URL }%
	\providecommand \Eprint [0]{\href }%
	\providecommand \doibase [0]{http://dx.doi.org/}%
	\providecommand \selectlanguage [0]{\@gobble}%
	\providecommand \bibinfo  [0]{\@secondoftwo}%
	\providecommand \bibfield  [0]{\@secondoftwo}%
	\providecommand \translation [1]{[#1]}%
	\providecommand \BibitemOpen [0]{}%
	\providecommand \bibitemStop [0]{}%
	\providecommand \bibitemNoStop [0]{.\EOS\space}%
	\providecommand \EOS [0]{\spacefactor3000\relax}%
	\providecommand \BibitemShut  [1]{\csname bibitem#1\endcsname}%
	\let\auto@bib@innerbib\@empty
	\bibitem [{\citenamefont {Boccaletti}\ \emph {et~al.}(2014)\citenamefont
		{Boccaletti}, \citenamefont {Bianconi}, \citenamefont {Criado}, \citenamefont
		{Del~Genio}, \citenamefont {G{\'o}mez-Gardenes}, \citenamefont {Romance},
		\citenamefont {Sendina-Nadal}, \citenamefont {Wang},\ and\ \citenamefont
		{Zanin}}]{boccaletti2014structure}%
	\BibitemOpen
	\bibfield  {author} {\bibinfo {author} {\bibfnamefont {S.}~\bibnamefont
			{Boccaletti}}, \bibinfo {author} {\bibfnamefont {G.}~\bibnamefont
			{Bianconi}}, \bibinfo {author} {\bibfnamefont {R.}~\bibnamefont {Criado}},
		\bibinfo {author} {\bibfnamefont {C.~I.}\ \bibnamefont {Del~Genio}}, \bibinfo
		{author} {\bibfnamefont {J.}~\bibnamefont {G{\'o}mez-Gardenes}}, \bibinfo
		{author} {\bibfnamefont {M.}~\bibnamefont {Romance}}, \bibinfo {author}
		{\bibfnamefont {I.}~\bibnamefont {Sendina-Nadal}}, \bibinfo {author}
		{\bibfnamefont {Z.}~\bibnamefont {Wang}}, \ and\ \bibinfo {author}
		{\bibfnamefont {M.}~\bibnamefont {Zanin}},\ }\href@noop {} {\bibfield
		{journal} {\bibinfo  {journal} {Physics Reports}\ }\textbf {\bibinfo {volume}
			{544}},\ \bibinfo {pages} {1} (\bibinfo {year} {2014})}\BibitemShut {NoStop}%
	\bibitem [{\citenamefont {Bianconi}(2018)}]{bianconi2018multilayer}%
	\BibitemOpen
	\bibfield  {author} {\bibinfo {author} {\bibfnamefont {G.}~\bibnamefont
			{Bianconi}},\ }\href@noop {} {\emph {\bibinfo {title} {Multilayer Networks:
				Structure and Function}}}\ (\bibinfo  {publisher} {Oxford University Press},\
	\bibinfo {year} {2018})\BibitemShut {NoStop}%
	\bibitem [{\citenamefont {Cardillo}\ \emph
		{et~al.}(2013{\natexlab{a}})\citenamefont {Cardillo}, \citenamefont
		{G{\'o}mez-Gardenes}, \citenamefont {Zanin}, \citenamefont {Romance},
		\citenamefont {Papo}, \citenamefont {Del~Pozo},\ and\ \citenamefont
		{Boccaletti}}]{mobility}%
	\BibitemOpen
	\bibfield  {author} {\bibinfo {author} {\bibfnamefont {A.}~\bibnamefont
			{Cardillo}}, \bibinfo {author} {\bibfnamefont {J.}~\bibnamefont
			{G{\'o}mez-Gardenes}}, \bibinfo {author} {\bibfnamefont {M.}~\bibnamefont
			{Zanin}}, \bibinfo {author} {\bibfnamefont {M.}~\bibnamefont {Romance}},
		\bibinfo {author} {\bibfnamefont {D.}~\bibnamefont {Papo}}, \bibinfo {author}
		{\bibfnamefont {F.}~\bibnamefont {Del~Pozo}}, \ and\ \bibinfo {author}
		{\bibfnamefont {S.}~\bibnamefont {Boccaletti}},\ }\href@noop {} {\bibfield
		{journal} {\bibinfo  {journal} {Scientific reports}\ }\textbf {\bibinfo
			{volume} {3}},\ \bibinfo {pages} {1} (\bibinfo {year}
		{2013}{\natexlab{a}})}\BibitemShut {NoStop}%
	\bibitem [{\citenamefont {Brummitt}\ \emph {et~al.}(2012)\citenamefont
		{Brummitt}, \citenamefont {D’Souza},\ and\ \citenamefont {Leicht}}]{power}%
	\BibitemOpen
	\bibfield  {author} {\bibinfo {author} {\bibfnamefont {C.~D.}\ \bibnamefont
			{Brummitt}}, \bibinfo {author} {\bibfnamefont {R.~M.}\ \bibnamefont
			{D’Souza}}, \ and\ \bibinfo {author} {\bibfnamefont {E.~A.}\ \bibnamefont
			{Leicht}},\ }\href@noop {} {\bibfield  {journal} {\bibinfo  {journal}
			{Proceedings of the national academy of sciences}\ }\textbf {\bibinfo
			{volume} {109}},\ \bibinfo {pages} {E680} (\bibinfo {year}
		{2012})}\BibitemShut {NoStop}%
	\bibitem [{\citenamefont {Cardillo}\ \emph
		{et~al.}(2013{\natexlab{b}})\citenamefont {Cardillo}, \citenamefont {Zanin},
		\citenamefont {G{\'o}mez-Gardenes}, \citenamefont {Romance}, \citenamefont
		{del Amo},\ and\ \citenamefont {Boccaletti}}]{air}%
	\BibitemOpen
	\bibfield  {author} {\bibinfo {author} {\bibfnamefont {A.}~\bibnamefont
			{Cardillo}}, \bibinfo {author} {\bibfnamefont {M.}~\bibnamefont {Zanin}},
		\bibinfo {author} {\bibfnamefont {J.}~\bibnamefont {G{\'o}mez-Gardenes}},
		\bibinfo {author} {\bibfnamefont {M.}~\bibnamefont {Romance}}, \bibinfo
		{author} {\bibfnamefont {A.~J.~G.}\ \bibnamefont {del Amo}}, \ and\ \bibinfo
		{author} {\bibfnamefont {S.}~\bibnamefont {Boccaletti}},\ }\href@noop {}
	{\bibfield  {journal} {\bibinfo  {journal} {The European Physical Journal
				Special Topics}\ }\textbf {\bibinfo {volume} {215}},\ \bibinfo {pages} {23}
		(\bibinfo {year} {2013}{\natexlab{b}})}\BibitemShut {NoStop}%
	\bibitem [{\citenamefont {Szell}\ \emph {et~al.}(2010)\citenamefont {Szell},
		\citenamefont {Lambiotte},\ and\ \citenamefont {Thurner}}]{social}%
	\BibitemOpen
	\bibfield  {author} {\bibinfo {author} {\bibfnamefont {M.}~\bibnamefont
			{Szell}}, \bibinfo {author} {\bibfnamefont {R.}~\bibnamefont {Lambiotte}}, \
		and\ \bibinfo {author} {\bibfnamefont {S.}~\bibnamefont {Thurner}},\
	}\href@noop {} {\bibfield  {journal} {\bibinfo  {journal} {Proceedings of the
				National Academy of Sciences}\ }\textbf {\bibinfo {volume} {107}},\ \bibinfo
		{pages} {13636} (\bibinfo {year} {2010})}\BibitemShut {NoStop}%
	\bibitem [{\citenamefont {Pilosof}\ \emph {et~al.}(2017)\citenamefont
		{Pilosof}, \citenamefont {Porter}, \citenamefont {Pascual},\ and\
		\citenamefont {K{\'e}fi}}]{ecology}%
	\BibitemOpen
	\bibfield  {author} {\bibinfo {author} {\bibfnamefont {S.}~\bibnamefont
			{Pilosof}}, \bibinfo {author} {\bibfnamefont {M.~A.}\ \bibnamefont {Porter}},
		\bibinfo {author} {\bibfnamefont {M.}~\bibnamefont {Pascual}}, \ and\
		\bibinfo {author} {\bibfnamefont {S.}~\bibnamefont {K{\'e}fi}},\ }\href@noop
	{} {\bibfield  {journal} {\bibinfo  {journal} {Nature Ecology \& Evolution}\
		}\textbf {\bibinfo {volume} {1}},\ \bibinfo {pages} {1} (\bibinfo {year}
		{2017})}\BibitemShut {NoStop}%
	\bibitem [{\citenamefont {Sorrentino}(2012)}]{sorrentino_njp}%
	\BibitemOpen
	\bibfield  {author} {\bibinfo {author} {\bibfnamefont {F.}~\bibnamefont
			{Sorrentino}},\ }\href@noop {} {\bibfield  {journal} {\bibinfo  {journal}
			{New Journal of Physics}\ }\textbf {\bibinfo {volume} {14}},\ \bibinfo
		{pages} {033035} (\bibinfo {year} {2012})}\BibitemShut {NoStop}%
	\bibitem [{\citenamefont {Sorrentino}\ and\ \citenamefont
		{Ott}(2007)}]{sorrentino2007network}%
	\BibitemOpen
	\bibfield  {author} {\bibinfo {author} {\bibfnamefont {F.}~\bibnamefont
			{Sorrentino}}\ and\ \bibinfo {author} {\bibfnamefont {E.}~\bibnamefont
			{Ott}},\ }\href@noop {} {\bibfield  {journal} {\bibinfo  {journal} {Physical
				Review E}\ }\textbf {\bibinfo {volume} {76}},\ \bibinfo {pages} {056114}
		(\bibinfo {year} {2007})}\BibitemShut {NoStop}%
	\bibitem [{\citenamefont {Ha}\ \emph {et~al.}(2017)\citenamefont {Ha},
		\citenamefont {Dai},\ and\ \citenamefont {Le}}]{DBLP:conf/iclr/HaDL17}%
	\BibitemOpen
	\bibfield  {author} {\bibinfo {author} {\bibfnamefont {D.}~\bibnamefont
			{Ha}}, \bibinfo {author} {\bibfnamefont {A.~M.}\ \bibnamefont {Dai}}, \ and\
		\bibinfo {author} {\bibfnamefont {Q.~V.}\ \bibnamefont {Le}},\ }in\
	\href@noop {} {\emph {\bibinfo {booktitle} {5th International Conference on
				Learning Representations, {ICLR} 2017, Toulon, France, April 24-26, 2017,
				Conference Track Proceedings}}}\ (\bibinfo  {publisher} {OpenReview.net},\
	\bibinfo {year} {2017})\BibitemShut {NoStop}%
	\bibitem [{\citenamefont {Rakshit}\ \emph
		{et~al.}(2018{\natexlab{a}})\citenamefont {Rakshit}, \citenamefont {Bera},
		\citenamefont {Ghosh},\ and\ \citenamefont {Sinha}}]{rakshit2018emergence}%
	\BibitemOpen
	\bibfield  {author} {\bibinfo {author} {\bibfnamefont {S.}~\bibnamefont
			{Rakshit}}, \bibinfo {author} {\bibfnamefont {B.~K.}\ \bibnamefont {Bera}},
		\bibinfo {author} {\bibfnamefont {D.}~\bibnamefont {Ghosh}}, \ and\ \bibinfo
		{author} {\bibfnamefont {S.}~\bibnamefont {Sinha}},\ }\href@noop {}
	{\bibfield  {journal} {\bibinfo  {journal} {Physical Review E}\ }\textbf
		{\bibinfo {volume} {97}},\ \bibinfo {pages} {052304} (\bibinfo {year}
		{2018}{\natexlab{a}})}\BibitemShut {NoStop}%
	\bibitem [{\citenamefont {Pikovsky}\ \emph {et~al.}(2001)\citenamefont
		{Pikovsky}, \citenamefont {Rosenblum},\ and\ \citenamefont
		{Kurths}}]{synchronization1}%
	\BibitemOpen
	\bibfield  {author} {\bibinfo {author} {\bibfnamefont {A.}~\bibnamefont
			{Pikovsky}}, \bibinfo {author} {\bibfnamefont {M.}~\bibnamefont {Rosenblum}},
		\ and\ \bibinfo {author} {\bibfnamefont {J.}~\bibnamefont {Kurths}},\
	}\href@noop {} {\emph {\bibinfo {title} {Synchronization: A Universal Concept
				in Nonlinear Sciences}}},\ Cambridge Nonlinear Science Series\ (\bibinfo
	{publisher} {Cambridge University Press},\ \bibinfo {year}
	{2001})\BibitemShut {NoStop}%
	\bibitem [{\citenamefont {Boccaletti}\ \emph {et~al.}(2002)\citenamefont
		{Boccaletti}, \citenamefont {Kurths}, \citenamefont {Osipov}, \citenamefont
		{Valladares},\ and\ \citenamefont {Zhou}}]{synchronization2}%
	\BibitemOpen
	\bibfield  {author} {\bibinfo {author} {\bibfnamefont {S.}~\bibnamefont
			{Boccaletti}}, \bibinfo {author} {\bibfnamefont {J.}~\bibnamefont {Kurths}},
		\bibinfo {author} {\bibfnamefont {G.}~\bibnamefont {Osipov}}, \bibinfo
		{author} {\bibfnamefont {D.}~\bibnamefont {Valladares}}, \ and\ \bibinfo
		{author} {\bibfnamefont {C.}~\bibnamefont {Zhou}},\ }\href@noop {} {\bibfield
		{journal} {\bibinfo  {journal} {Physics Reports}\ }\textbf {\bibinfo
			{volume} {366}},\ \bibinfo {pages} {1} (\bibinfo {year} {2002})}\BibitemShut
	{NoStop}%
	\bibitem [{\citenamefont {Arenas}\ \emph {et~al.}(2008)\citenamefont {Arenas},
		\citenamefont {D{\'\i}az-Guilera}, \citenamefont {Kurths}, \citenamefont
		{Moreno},\ and\ \citenamefont {Zhou}}]{synchronization3}%
	\BibitemOpen
	\bibfield  {author} {\bibinfo {author} {\bibfnamefont {A.}~\bibnamefont
			{Arenas}}, \bibinfo {author} {\bibfnamefont {A.}~\bibnamefont
			{D{\'\i}az-Guilera}}, \bibinfo {author} {\bibfnamefont {J.}~\bibnamefont
			{Kurths}}, \bibinfo {author} {\bibfnamefont {Y.}~\bibnamefont {Moreno}}, \
		and\ \bibinfo {author} {\bibfnamefont {C.}~\bibnamefont {Zhou}},\ }\href@noop
	{} {\bibfield  {journal} {\bibinfo  {journal} {Physics Reports}\ }\textbf
		{\bibinfo {volume} {469}},\ \bibinfo {pages} {93} (\bibinfo {year}
		{2008})}\BibitemShut {NoStop}%
	\bibitem [{\citenamefont {G{\'o}mez-Gardenes}\ \emph
		{et~al.}(2007)\citenamefont {G{\'o}mez-Gardenes}, \citenamefont {Moreno},\
		and\ \citenamefont {Arenas}}]{synchronization4}%
	\BibitemOpen
	\bibfield  {author} {\bibinfo {author} {\bibfnamefont {J.}~\bibnamefont
			{G{\'o}mez-Gardenes}}, \bibinfo {author} {\bibfnamefont {Y.}~\bibnamefont
			{Moreno}}, \ and\ \bibinfo {author} {\bibfnamefont {A.}~\bibnamefont
			{Arenas}},\ }\href@noop {} {\bibfield  {journal} {\bibinfo  {journal}
			{Physical Review Letters}\ }\textbf {\bibinfo {volume} {98}},\ \bibinfo
		{pages} {034101} (\bibinfo {year} {2007})}\BibitemShut {NoStop}%
	\bibitem [{\citenamefont {Skardal}\ \emph {et~al.}(2014)\citenamefont
		{Skardal}, \citenamefont {Taylor},\ and\ \citenamefont
		{Sun}}]{skardal2014optimal}%
	\BibitemOpen
	\bibfield  {author} {\bibinfo {author} {\bibfnamefont {P.~S.}\ \bibnamefont
			{Skardal}}, \bibinfo {author} {\bibfnamefont {D.}~\bibnamefont {Taylor}}, \
		and\ \bibinfo {author} {\bibfnamefont {J.}~\bibnamefont {Sun}},\ }\href@noop
	{} {\bibfield  {journal} {\bibinfo  {journal} {Physical Review Letters}\
		}\textbf {\bibinfo {volume} {113}},\ \bibinfo {pages} {144101} (\bibinfo
		{year} {2014})}\BibitemShut {NoStop}%
	\bibitem [{\citenamefont {Restrepo}\ \emph {et~al.}(2006)\citenamefont
		{Restrepo}, \citenamefont {Ott},\ and\ \citenamefont
		{Hunt}}]{restrepo2006synchronization}%
	\BibitemOpen
	\bibfield  {author} {\bibinfo {author} {\bibfnamefont {J.~G.}\ \bibnamefont
			{Restrepo}}, \bibinfo {author} {\bibfnamefont {E.}~\bibnamefont {Ott}}, \
		and\ \bibinfo {author} {\bibfnamefont {B.~R.}\ \bibnamefont {Hunt}},\
	}\href@noop {} {\bibfield  {journal} {\bibinfo  {journal} {Chaos: An
				Interdisciplinary Journal of Nonlinear Science}\ }\textbf {\bibinfo {volume}
			{16}},\ \bibinfo {pages} {015107} (\bibinfo {year} {2006})}\BibitemShut
	{NoStop}%
	\bibitem [{\citenamefont {Restrepo}\ \emph {et~al.}(2005)\citenamefont
		{Restrepo}, \citenamefont {Ott},\ and\ \citenamefont
		{Hunt}}]{restrepo2005onset}%
	\BibitemOpen
	\bibfield  {author} {\bibinfo {author} {\bibfnamefont {J.~G.}\ \bibnamefont
			{Restrepo}}, \bibinfo {author} {\bibfnamefont {E.}~\bibnamefont {Ott}}, \
		and\ \bibinfo {author} {\bibfnamefont {B.~R.}\ \bibnamefont {Hunt}},\
	}\href@noop {} {\bibfield  {journal} {\bibinfo  {journal} {Physical Review
				E}\ }\textbf {\bibinfo {volume} {71}},\ \bibinfo {pages} {036151} (\bibinfo
		{year} {2005})}\BibitemShut {NoStop}%
	\bibitem [{\citenamefont {Pecora}\ and\ \citenamefont
		{Carroll}(1998)}]{pecora1998master}%
	\BibitemOpen
	\bibfield  {author} {\bibinfo {author} {\bibfnamefont {L.~M.}\ \bibnamefont
			{Pecora}}\ and\ \bibinfo {author} {\bibfnamefont {T.~L.}\ \bibnamefont
			{Carroll}},\ }\href@noop {} {\bibfield  {journal} {\bibinfo  {journal}
			{Physical Review Letters}\ }\textbf {\bibinfo {volume} {80}},\ \bibinfo
		{pages} {2109} (\bibinfo {year} {1998})}\BibitemShut {NoStop}%
	\bibitem [{\citenamefont {Sun}\ \emph {et~al.}(2009)\citenamefont {Sun},
		\citenamefont {Bollt},\ and\ \citenamefont {Nishikawa}}]{sun2009master}%
	\BibitemOpen
	\bibfield  {author} {\bibinfo {author} {\bibfnamefont {J.}~\bibnamefont
			{Sun}}, \bibinfo {author} {\bibfnamefont {E.~M.}\ \bibnamefont {Bollt}}, \
		and\ \bibinfo {author} {\bibfnamefont {T.}~\bibnamefont {Nishikawa}},\
	}\href@noop {} {\bibfield  {journal} {\bibinfo  {journal} {EPL (Europhysics
				Letters)}\ }\textbf {\bibinfo {volume} {85}},\ \bibinfo {pages} {60011}
		(\bibinfo {year} {2009})}\BibitemShut {NoStop}%
	\bibitem [{\citenamefont {Nishikawa}\ and\ \citenamefont
		{Motter}(2006)}]{nishikawa2006synchronization}%
	\BibitemOpen
	\bibfield  {author} {\bibinfo {author} {\bibfnamefont {T.}~\bibnamefont
			{Nishikawa}}\ and\ \bibinfo {author} {\bibfnamefont {A.~E.}\ \bibnamefont
			{Motter}},\ }\href@noop {} {\bibfield  {journal} {\bibinfo  {journal}
			{Physical Review E}\ }\textbf {\bibinfo {volume} {73}},\ \bibinfo {pages}
		{065106} (\bibinfo {year} {2006})}\BibitemShut {NoStop}%
	\bibitem [{\citenamefont {Irving}\ and\ \citenamefont
		{Sorrentino}(2012)}]{sbd_sorrentino}%
	\BibitemOpen
	\bibfield  {author} {\bibinfo {author} {\bibfnamefont {D.}~\bibnamefont
			{Irving}}\ and\ \bibinfo {author} {\bibfnamefont {F.}~\bibnamefont
			{Sorrentino}},\ }\href@noop {} {\bibfield  {journal} {\bibinfo  {journal}
			{Physical Review E}\ }\textbf {\bibinfo {volume} {86}},\ \bibinfo {pages}
		{056102} (\bibinfo {year} {2012})}\BibitemShut {NoStop}%
	\bibitem [{\citenamefont {Belykh}\ \emph {et~al.}(2019)\citenamefont {Belykh},
		\citenamefont {Carter},\ and\ \citenamefont
		{Jeter}}]{belykh2019synchronization}%
	\BibitemOpen
	\bibfield  {author} {\bibinfo {author} {\bibfnamefont {I.}~\bibnamefont
			{Belykh}}, \bibinfo {author} {\bibfnamefont {D.}~\bibnamefont {Carter}}, \
		and\ \bibinfo {author} {\bibfnamefont {R.}~\bibnamefont {Jeter}},\
	}\href@noop {} {\bibfield  {journal} {\bibinfo  {journal} {SIAM Journal on
				Applied Dynamical Systems}\ }\textbf {\bibinfo {volume} {18}},\ \bibinfo
		{pages} {2267} (\bibinfo {year} {2019})}\BibitemShut {NoStop}%
	\bibitem [{\citenamefont {Berner}\ \emph {et~al.}(2021)\citenamefont {Berner},
		\citenamefont {Mehrmann}, \citenamefont {Schöll},\ and\ \citenamefont
		{Yanchuk}}]{multiplex_decom}%
	\BibitemOpen
	\bibfield  {author} {\bibinfo {author} {\bibfnamefont {R.}~\bibnamefont
			{Berner}}, \bibinfo {author} {\bibfnamefont {V.}~\bibnamefont {Mehrmann}},
		\bibinfo {author} {\bibfnamefont {E.}~\bibnamefont {Schöll}}, \ and\
		\bibinfo {author} {\bibfnamefont {S.}~\bibnamefont {Yanchuk}},\ }\href@noop
	{} {\bibfield  {journal} {\bibinfo  {journal} {SIAM Journal on Applied
				Dynamical Systems}\ }\textbf {\bibinfo {volume} {20}},\ \bibinfo {pages}
		{1752} (\bibinfo {year} {2021})}\BibitemShut {NoStop}%
	\bibitem [{\citenamefont {Rakshit}\ \emph {et~al.}(2022)\citenamefont
		{Rakshit}, \citenamefont {Parastesh}, \citenamefont {Chowdhury},
		\citenamefont {Jafari}, \citenamefont {Kurths},\ and\ \citenamefont
		{Ghosh}}]{rakshit2021relay}%
	\BibitemOpen
	\bibfield  {author} {\bibinfo {author} {\bibfnamefont {S.}~\bibnamefont
			{Rakshit}}, \bibinfo {author} {\bibfnamefont {F.}~\bibnamefont {Parastesh}},
		\bibinfo {author} {\bibfnamefont {S.~N.}\ \bibnamefont {Chowdhury}}, \bibinfo
		{author} {\bibfnamefont {S.}~\bibnamefont {Jafari}}, \bibinfo {author}
		{\bibfnamefont {J.}~\bibnamefont {Kurths}}, \ and\ \bibinfo {author}
		{\bibfnamefont {D.}~\bibnamefont {Ghosh}},\ }\href@noop {} {\bibfield
		{journal} {\bibinfo  {journal} {Nonlinearity}\ }\textbf {\bibinfo {volume}
			{35}},\ \bibinfo {pages} {681} (\bibinfo {year} {2022})}\BibitemShut
	{NoStop}%
	\bibitem [{\citenamefont {Onnela}\ \emph {et~al.}(2007)\citenamefont {Onnela},
		\citenamefont {Saram{\"a}ki}, \citenamefont {Hyv{\"o}nen}, \citenamefont
		{Szab{\'o}}, \citenamefont {Lazer}, \citenamefont {Kaski}, \citenamefont
		{Kert{\'e}sz},\ and\ \citenamefont {Barab{\'a}si}}]{onnela2007structure}%
	\BibitemOpen
	\bibfield  {author} {\bibinfo {author} {\bibfnamefont {J.-P.}\ \bibnamefont
			{Onnela}}, \bibinfo {author} {\bibfnamefont {J.}~\bibnamefont
			{Saram{\"a}ki}}, \bibinfo {author} {\bibfnamefont {J.}~\bibnamefont
			{Hyv{\"o}nen}}, \bibinfo {author} {\bibfnamefont {G.}~\bibnamefont
			{Szab{\'o}}}, \bibinfo {author} {\bibfnamefont {D.}~\bibnamefont {Lazer}},
		\bibinfo {author} {\bibfnamefont {K.}~\bibnamefont {Kaski}}, \bibinfo
		{author} {\bibfnamefont {J.}~\bibnamefont {Kert{\'e}sz}}, \ and\ \bibinfo
		{author} {\bibfnamefont {A.-L.}\ \bibnamefont {Barab{\'a}si}},\ }\href@noop
	{} {\bibfield  {journal} {\bibinfo  {journal} {Proceedings of the national
				academy of sciences}\ }\textbf {\bibinfo {volume} {104}},\ \bibinfo {pages}
		{7332} (\bibinfo {year} {2007})}\BibitemShut {NoStop}%
	\bibitem [{\citenamefont {Pastor-Satorras}\ and\ \citenamefont
		{Vespignani}(2004)}]{pastor2004evolution}%
	\BibitemOpen
	\bibfield  {author} {\bibinfo {author} {\bibfnamefont {R.}~\bibnamefont
			{Pastor-Satorras}}\ and\ \bibinfo {author} {\bibfnamefont {A.}~\bibnamefont
			{Vespignani}},\ }\href@noop {} {\emph {\bibinfo {title} {Evolution and
				Structure of the Internet: A Statistical Physics Approach}}}\ (\bibinfo
	{publisher} {Cambridge University Press},\ \bibinfo {year}
	{2004})\BibitemShut {NoStop}%
	\bibitem [{\citenamefont {Skufca}\ and\ \citenamefont
		{Bollt}(2004)}]{skufca2004communication}%
	\BibitemOpen
	\bibfield  {author} {\bibinfo {author} {\bibfnamefont {J.~D.}\ \bibnamefont
			{Skufca}}\ and\ \bibinfo {author} {\bibfnamefont {E.~M.}\ \bibnamefont
			{Bollt}},\ }\href@noop {} {\bibfield  {journal} {\bibinfo  {journal}
			{Mathematical Biosciences \& Engineering}\ }\textbf {\bibinfo {volume} {1}},\
		\bibinfo {pages} {347} (\bibinfo {year} {2004})}\BibitemShut {NoStop}%
	\bibitem [{\citenamefont {Wasserman}\ \emph {et~al.}(1994)\citenamefont
		{Wasserman}, \citenamefont {Faust} \emph {et~al.}}]{wasserman1994social}%
	\BibitemOpen
	\bibfield  {author} {\bibinfo {author} {\bibfnamefont {S.}~\bibnamefont
			{Wasserman}}, \bibinfo {author} {\bibfnamefont {K.}~\bibnamefont {Faust}},
		\emph {et~al.},\ }\href@noop {} {\emph {\bibinfo {title} {Social Network
				Analysis: Methods and Applications}}}\ (\bibinfo  {publisher} {Cambridge
		University Press},\ \bibinfo {year} {1994})\BibitemShut {NoStop}%
	\bibitem [{\citenamefont {Motter}\ \emph {et~al.}(2013)\citenamefont {Motter},
		\citenamefont {Myers}, \citenamefont {Anghel},\ and\ \citenamefont
		{Nishikawa}}]{motter2013spontaneous}%
	\BibitemOpen
	\bibfield  {author} {\bibinfo {author} {\bibfnamefont {A.~E.}\ \bibnamefont
			{Motter}}, \bibinfo {author} {\bibfnamefont {S.~A.}\ \bibnamefont {Myers}},
		\bibinfo {author} {\bibfnamefont {M.}~\bibnamefont {Anghel}}, \ and\ \bibinfo
		{author} {\bibfnamefont {T.}~\bibnamefont {Nishikawa}},\ }\href@noop {}
	{\bibfield  {journal} {\bibinfo  {journal} {Nature Physics}\ }\textbf
		{\bibinfo {volume} {9}},\ \bibinfo {pages} {191} (\bibinfo {year}
		{2013})}\BibitemShut {NoStop}%
	\bibitem [{\citenamefont {Ghosh}\ \emph {et~al.}(2022)\citenamefont {Ghosh},
		\citenamefont {Frasca}, \citenamefont {Rizzo}, \citenamefont {Majhi},
		\citenamefont {Rakshit}, \citenamefont {Alfaro-Bittner},\ and\ \citenamefont
		{Boccaletti}}]{ghosh2022synchronized}%
	\BibitemOpen
	\bibfield  {author} {\bibinfo {author} {\bibfnamefont {D.}~\bibnamefont
			{Ghosh}}, \bibinfo {author} {\bibfnamefont {M.}~\bibnamefont {Frasca}},
		\bibinfo {author} {\bibfnamefont {A.}~\bibnamefont {Rizzo}}, \bibinfo
		{author} {\bibfnamefont {S.}~\bibnamefont {Majhi}}, \bibinfo {author}
		{\bibfnamefont {S.}~\bibnamefont {Rakshit}}, \bibinfo {author} {\bibfnamefont
			{K.}~\bibnamefont {Alfaro-Bittner}}, \ and\ \bibinfo {author} {\bibfnamefont
			{S.}~\bibnamefont {Boccaletti}},\ }\href@noop {} {\bibfield  {journal}
		{\bibinfo  {journal} {Physics Reports}\ }\textbf {\bibinfo {volume} {949}},\
		\bibinfo {pages} {1} (\bibinfo {year} {2022})}\BibitemShut {NoStop}%
	\bibitem [{\citenamefont {Holme}\ and\ \citenamefont
		{Saram{\"a}ki}(2012)}]{holme2012temporal}%
	\BibitemOpen
	\bibfield  {author} {\bibinfo {author} {\bibfnamefont {P.}~\bibnamefont
			{Holme}}\ and\ \bibinfo {author} {\bibfnamefont {J.}~\bibnamefont
			{Saram{\"a}ki}},\ }\href@noop {} {\bibfield  {journal} {\bibinfo  {journal}
			{Physics Reports}\ }\textbf {\bibinfo {volume} {519}},\ \bibinfo {pages} {97}
		(\bibinfo {year} {2012})}\BibitemShut {NoStop}%
	\bibitem [{\citenamefont {Taylor}\ \emph {et~al.}(2010)\citenamefont {Taylor},
		\citenamefont {Ott},\ and\ \citenamefont {Restrepo}}]{taylor2010spontaneous}%
	\BibitemOpen
	\bibfield  {author} {\bibinfo {author} {\bibfnamefont {D.}~\bibnamefont
			{Taylor}}, \bibinfo {author} {\bibfnamefont {E.}~\bibnamefont {Ott}}, \ and\
		\bibinfo {author} {\bibfnamefont {J.~G.}\ \bibnamefont {Restrepo}},\
	}\href@noop {} {\bibfield  {journal} {\bibinfo  {journal} {Physical Review
				E}\ }\textbf {\bibinfo {volume} {81}},\ \bibinfo {pages} {046214} (\bibinfo
		{year} {2010})}\BibitemShut {NoStop}%
	\bibitem [{\citenamefont {Stilwell}\ \emph {et~al.}(2006)\citenamefont
		{Stilwell}, \citenamefont {Bollt},\ and\ \citenamefont
		{Roberson}}]{stilwell2006sufficient}%
	\BibitemOpen
	\bibfield  {author} {\bibinfo {author} {\bibfnamefont {D.~J.}\ \bibnamefont
			{Stilwell}}, \bibinfo {author} {\bibfnamefont {E.~M.}\ \bibnamefont {Bollt}},
		\ and\ \bibinfo {author} {\bibfnamefont {D.~G.}\ \bibnamefont {Roberson}},\
	}\href@noop {} {\bibfield  {journal} {\bibinfo  {journal} {SIAM Journal on
				Applied Dynamical Systems}\ }\textbf {\bibinfo {volume} {5}},\ \bibinfo
		{pages} {140} (\bibinfo {year} {2006})}\BibitemShut {NoStop}%
	\bibitem [{\citenamefont {Belykh}\ \emph
		{et~al.}(2004{\natexlab{a}})\citenamefont {Belykh}, \citenamefont {Belykh},\
		and\ \citenamefont {Hasler}}]{belykh2004blinking}%
	\BibitemOpen
	\bibfield  {author} {\bibinfo {author} {\bibfnamefont {I.~V.}\ \bibnamefont
			{Belykh}}, \bibinfo {author} {\bibfnamefont {V.~N.}\ \bibnamefont {Belykh}},
		\ and\ \bibinfo {author} {\bibfnamefont {M.}~\bibnamefont {Hasler}},\
	}\href@noop {} {\bibfield  {journal} {\bibinfo  {journal} {Physica D:
				Nonlinear Phenomena}\ }\textbf {\bibinfo {volume} {195}},\ \bibinfo {pages}
		{188} (\bibinfo {year} {2004}{\natexlab{a}})}\BibitemShut {NoStop}%
	\bibitem [{\citenamefont {Belykh}\ \emph
		{et~al.}(2004{\natexlab{b}})\citenamefont {Belykh}, \citenamefont {Belykh},\
		and\ \citenamefont {Hasler}}]{belykh2004connection}%
	\BibitemOpen
	\bibfield  {author} {\bibinfo {author} {\bibfnamefont {V.~N.}\ \bibnamefont
			{Belykh}}, \bibinfo {author} {\bibfnamefont {I.~V.}\ \bibnamefont {Belykh}},
		\ and\ \bibinfo {author} {\bibfnamefont {M.}~\bibnamefont {Hasler}},\
	}\href@noop {} {\bibfield  {journal} {\bibinfo  {journal} {Physica D:
				nonlinear phenomena}\ }\textbf {\bibinfo {volume} {195}},\ \bibinfo {pages}
		{159} (\bibinfo {year} {2004}{\natexlab{b}})}\BibitemShut {NoStop}%
	\bibitem [{\citenamefont {Zhang}\ and\ \citenamefont
		{Motter}(2020)}]{zhang2020symmetry}%
	\BibitemOpen
	\bibfield  {author} {\bibinfo {author} {\bibfnamefont {Y.}~\bibnamefont
			{Zhang}}\ and\ \bibinfo {author} {\bibfnamefont {A.~E.}\ \bibnamefont
			{Motter}},\ }\href@noop {} {\bibfield  {journal} {\bibinfo  {journal} {SIAM
				Review}\ }\textbf {\bibinfo {volume} {62}},\ \bibinfo {pages} {817} (\bibinfo
		{year} {2020})}\BibitemShut {NoStop}%
	\bibitem [{\citenamefont {Zhang}\ \emph {et~al.}(2021)\citenamefont {Zhang},
		\citenamefont {Latora},\ and\ \citenamefont {Motter}}]{zhang2021unified}%
	\BibitemOpen
	\bibfield  {author} {\bibinfo {author} {\bibfnamefont {Y.}~\bibnamefont
			{Zhang}}, \bibinfo {author} {\bibfnamefont {V.}~\bibnamefont {Latora}}, \
		and\ \bibinfo {author} {\bibfnamefont {A.~E.}\ \bibnamefont {Motter}},\
	}\href@noop {} {\bibfield  {journal} {\bibinfo  {journal} {Communications
				Physics}\ }\textbf {\bibinfo {volume} {4}},\ \bibinfo {pages} {1} (\bibinfo
		{year} {2021})}\BibitemShut {NoStop}%
	\bibitem [{\citenamefont {Gambuzza}\ \emph {et~al.}(2015)\citenamefont
		{Gambuzza}, \citenamefont {Frasca},\ and\ \citenamefont
		{Gomez-Gardenes}}]{intra1}%
	\BibitemOpen
	\bibfield  {author} {\bibinfo {author} {\bibfnamefont {L.~V.}\ \bibnamefont
			{Gambuzza}}, \bibinfo {author} {\bibfnamefont {M.}~\bibnamefont {Frasca}}, \
		and\ \bibinfo {author} {\bibfnamefont {J.}~\bibnamefont {Gomez-Gardenes}},\
	}\href@noop {} {\bibfield  {journal} {\bibinfo  {journal} {EPL (Europhysics
				Letters)}\ }\textbf {\bibinfo {volume} {110}},\ \bibinfo {pages} {20010}
		(\bibinfo {year} {2015})}\BibitemShut {NoStop}%
	\bibitem [{\citenamefont {Rakshit}\ \emph {et~al.}(2020)\citenamefont
		{Rakshit}, \citenamefont {Bera}, \citenamefont {Bollt},\ and\ \citenamefont
		{Ghosh}}]{rakshit2020intralayer}%
	\BibitemOpen
	\bibfield  {author} {\bibinfo {author} {\bibfnamefont {S.}~\bibnamefont
			{Rakshit}}, \bibinfo {author} {\bibfnamefont {B.~K.}\ \bibnamefont {Bera}},
		\bibinfo {author} {\bibfnamefont {E.~M.}\ \bibnamefont {Bollt}}, \ and\
		\bibinfo {author} {\bibfnamefont {D.}~\bibnamefont {Ghosh}},\ }\href@noop {}
	{\bibfield  {journal} {\bibinfo  {journal} {SIAM Journal on Applied Dynamical
				Systems}\ }\textbf {\bibinfo {volume} {19}},\ \bibinfo {pages} {918}
		(\bibinfo {year} {2020})}\BibitemShut {NoStop}%
	\bibitem [{\citenamefont {Rakshit}\ \emph
		{et~al.}(2018{\natexlab{b}})\citenamefont {Rakshit}, \citenamefont {Bera},\
		and\ \citenamefont {Ghosh}}]{rakshit2018synchronization}%
	\BibitemOpen
	\bibfield  {author} {\bibinfo {author} {\bibfnamefont {S.}~\bibnamefont
			{Rakshit}}, \bibinfo {author} {\bibfnamefont {B.~K.}\ \bibnamefont {Bera}}, \
		and\ \bibinfo {author} {\bibfnamefont {D.}~\bibnamefont {Ghosh}},\
	}\href@noop {} {\bibfield  {journal} {\bibinfo  {journal} {Physical Review
				E}\ }\textbf {\bibinfo {volume} {98}},\ \bibinfo {pages} {032305} (\bibinfo
		{year} {2018}{\natexlab{b}})}\BibitemShut {NoStop}%
	\bibitem [{\citenamefont {Rakshit}\ \emph {et~al.}(2017)\citenamefont
		{Rakshit}, \citenamefont {Majhi}, \citenamefont {Bera}, \citenamefont
		{Sinha},\ and\ \citenamefont {Ghosh}}]{rakshit2017time}%
	\BibitemOpen
	\bibfield  {author} {\bibinfo {author} {\bibfnamefont {S.}~\bibnamefont
			{Rakshit}}, \bibinfo {author} {\bibfnamefont {S.}~\bibnamefont {Majhi}},
		\bibinfo {author} {\bibfnamefont {B.~K.}\ \bibnamefont {Bera}}, \bibinfo
		{author} {\bibfnamefont {S.}~\bibnamefont {Sinha}}, \ and\ \bibinfo {author}
		{\bibfnamefont {D.}~\bibnamefont {Ghosh}},\ }\href@noop {} {\bibfield
		{journal} {\bibinfo  {journal} {Physical Review E}\ }\textbf {\bibinfo
			{volume} {96}},\ \bibinfo {pages} {062308} (\bibinfo {year}
		{2017})}\BibitemShut {NoStop}%
	\bibitem [{\citenamefont {Maehara}\ and\ \citenamefont
		{Murota}(2011)}]{maehara2011algorithm}%
	\BibitemOpen
	\bibfield  {author} {\bibinfo {author} {\bibfnamefont {T.}~\bibnamefont
			{Maehara}}\ and\ \bibinfo {author} {\bibfnamefont {K.}~\bibnamefont
			{Murota}},\ }\href@noop {} {\bibfield  {journal} {\bibinfo  {journal} {SIAM
				Journal on Matrix Analysis and Applications}\ }\textbf {\bibinfo {volume}
			{32}},\ \bibinfo {pages} {605} (\bibinfo {year} {2011})}\BibitemShut
	{NoStop}%
	\bibitem [{\citenamefont {Strogatz}(2018)}]{strogatz2018nonlinear}%
	\BibitemOpen
	\bibfield  {author} {\bibinfo {author} {\bibfnamefont {S.~H.}\ \bibnamefont
			{Strogatz}},\ }\href@noop {} {\emph {\bibinfo {title} {Nonlinear Dynamics and
				Chaos with Student Solutions Manual: With Applications to Physics, Biology,
				Chemistry, and Engineering}}}\ (\bibinfo  {publisher} {CRC press},\ \bibinfo
	{year} {2018})\BibitemShut {NoStop}%
	\bibitem [{\citenamefont {R{\"o}ssler}(1976)}]{rossler1976equation}%
	\BibitemOpen
	\bibfield  {author} {\bibinfo {author} {\bibfnamefont {O.~E.}\ \bibnamefont
			{R{\"o}ssler}},\ }\href@noop {} {\bibfield  {journal} {\bibinfo  {journal}
			{Physics Letters A}\ }\textbf {\bibinfo {volume} {57}},\ \bibinfo {pages}
		{397} (\bibinfo {year} {1976})}\BibitemShut {NoStop}%
	\bibitem [{\citenamefont {Sherman}(1994)}]{sherman1994anti}%
	\BibitemOpen
	\bibfield  {author} {\bibinfo {author} {\bibfnamefont {A.}~\bibnamefont
			{Sherman}},\ }\href@noop {} {\bibfield  {journal} {\bibinfo  {journal}
			{Bulletin of mathematical biology}\ }\textbf {\bibinfo {volume} {56}},\
		\bibinfo {pages} {811} (\bibinfo {year} {1994})}\BibitemShut {NoStop}%
	\bibitem [{\citenamefont {Jalil}\ \emph {et~al.}(2012)\citenamefont {Jalil},
		\citenamefont {Belykh},\ and\ \citenamefont {Shilnikov}}]{jalil2012spikes}%
	\BibitemOpen
	\bibfield  {author} {\bibinfo {author} {\bibfnamefont {S.}~\bibnamefont
			{Jalil}}, \bibinfo {author} {\bibfnamefont {I.}~\bibnamefont {Belykh}}, \
		and\ \bibinfo {author} {\bibfnamefont {A.}~\bibnamefont {Shilnikov}},\
	}\href@noop {} {\bibfield  {journal} {\bibinfo  {journal} {Physical Review
				E}\ }\textbf {\bibinfo {volume} {85}},\ \bibinfo {pages} {036214} (\bibinfo
		{year} {2012})}\BibitemShut {NoStop}%
	\bibitem [{\citenamefont {Erd{\"o}s}\ and\ \citenamefont
		{R{\'e}nyi}(2011)}]{erdos2011evolution}%
	\BibitemOpen
	\bibfield  {author} {\bibinfo {author} {\bibfnamefont {P.}~\bibnamefont
			{Erd{\"o}s}}\ and\ \bibinfo {author} {\bibfnamefont {A.}~\bibnamefont
			{R{\'e}nyi}},\ }in\ \href@noop {} {\emph {\bibinfo {booktitle} {The Structure
				and Dynamics of Networks}}}\ (\bibinfo  {publisher} {Princeton University
		Press},\ \bibinfo {year} {2011})\ pp.\ \bibinfo {pages} {38--82}\BibitemShut
	{NoStop}%
	\bibitem [{\citenamefont {Watts}\ and\ \citenamefont
		{Strogatz}(1998)}]{watts1998collective}%
	\BibitemOpen
	\bibfield  {author} {\bibinfo {author} {\bibfnamefont {D.~J.}\ \bibnamefont
			{Watts}}\ and\ \bibinfo {author} {\bibfnamefont {S.~H.}\ \bibnamefont
			{Strogatz}},\ }\href@noop {} {\bibfield  {journal} {\bibinfo  {journal}
			{Nature}\ }\textbf {\bibinfo {volume} {393}},\ \bibinfo {pages} {440}
		(\bibinfo {year} {1998})}\BibitemShut {NoStop}%
	\bibitem [{\citenamefont {Sporns}(2010)}]{sporns2010networks}%
	\BibitemOpen
	\bibfield  {author} {\bibinfo {author} {\bibfnamefont {O.}~\bibnamefont
			{Sporns}},\ }\href@noop {} {\emph {\bibinfo {title} {Networks of the
				Brain}}}\ (\bibinfo  {publisher} {MIT press},\ \bibinfo {year}
	{2010})\BibitemShut {NoStop}%
	\bibitem [{\citenamefont {Bera}\ \emph {et~al.}(2019)\citenamefont {Bera},
		\citenamefont {Rakshit},\ and\ \citenamefont {Ghosh}}]{bera2019intralayer}%
	\BibitemOpen
	\bibfield  {author} {\bibinfo {author} {\bibfnamefont {B.~K.}\ \bibnamefont
			{Bera}}, \bibinfo {author} {\bibfnamefont {S.}~\bibnamefont {Rakshit}}, \
		and\ \bibinfo {author} {\bibfnamefont {D.}~\bibnamefont {Ghosh}},\
	}\href@noop {} {\bibfield  {journal} {\bibinfo  {journal} {The European
				Physical Journal Special Topics}\ }\textbf {\bibinfo {volume} {228}},\
		\bibinfo {pages} {2441} (\bibinfo {year} {2019})}\BibitemShut {NoStop}%
\end{thebibliography}
%

\end{document}